\def\lesssim{\mathrel{\hbox{\rlap{\hbox{\lower4pt\hbox{$\sim$}}}\hbox{$<$}}}}
\def\gtrsim{\mathrel{\hbox{\rlap{\hbox{\lower4pt\hbox{$\sim$}}}\hbox{$>$}}}}

\documentclass[natbib]{svjour3}                     
\smartqed  
\usepackage{graphicx}
 \usepackage{aps-bibstyle}  
\begin{document}

\def\bfnote #1]{{\bf #1]}}

\title{Some glimpses from helioseismology at the dynamics of the deep solar interior}


\titlerunning{Glimpes at deep solar dynamics}        

\author{D.O. Gough
}


\institute{Institute of Astronomy
\& Department of Applied Mathematics and Theoretical Physics,
Madingley Road, Cambridge, CB3 0HA, UK
}

\date{Received: date / Accepted: date}

\maketitle

\begin{abstract}
Helioseismology has taught us a great deal about the stratification and kinematics of the solar interior, sufficient for us to embark upon dynamical studies more detailed than have been possible before.  The most 
sophisticated studies to date have been the very impressive numerical simulations of the convection zone, from which, especially in recent years, a great deal has been learnt.  Those simulations, and the seismological evidence with which they are being confronted, are reviewed elsewhere in this volume.   Our understanding of the global dynamics of the radiative interior of the Sun is in a much more primitive state.  Nevertheless, some progress has been made, and seismological inference has provided us with evidence of more to come.  Some of that I summarize here, mentioning in passing hints that are pointing the way to the future.  

\keywords{Sun \and helioseismology \and dynamics}
\end{abstract}

\section{Introduction}
\label{sec:1}

%

The most studied dynamics of the Sun occurs in the convective envelope, where the timescales are human. 
The large-scale convective flow and its associated angular-momentum transport is to some degree accessible to seismological probing.  However, I shall pay little attention to that here,  because it is reviewed elsewhere in this volume by \citet{brunetalISSI2015SSRv..tmp..100B} and \citet{ hanasogeetalISSI2013SSRv..tmp..100B}, 
who discuss principally the results of the impressive modern simulations and the physics that has been gleaned from them. Instead, I shall try to reflect on the global behaviour of the Sun, recalling some of the issues that have concerned heliophysicists in the past -- issues that to some degree have been thought by commentators not in the field to have been resolved, yet in the light of our increasingly sophisticated helioseismological findings have left some room for more than a modicum of uncertainty. 

In order to establish the what-one-might-call standard view of the Sun, I shall first describe briefly what we believe to have been the principal dynamical processes that have influenced the evolution of the Sun to its present state. Only then do we have a basis for discussing the extent to which those beliefs should really be relied upon. Quite naturally, I first adopt the simplest of descriptions, consistent with our general understanding of stellar evolution;  that seems to provide a gratifyingly accurate first-order class of models. There have been attempts to standardize some versions of those models, principally by John Bahcall. However, a universal standard has never been produced, because heliophysicists have not agreed on the manner in which the physics should be simplified. The discord has been aired many times, such as during IAU Colloquium 121 \citep{insidethesun1990ASSL..159..B}, at which Evry Schatzman \citep[reported by][]{dogopenquestions_1990ASSL..159..451G} favoured the inclusion,  by some means or other, of all the pertinent generally accepted physics, including macroscopic dynamics (represented in some precisely stated, albeit approximate, manner), in order to obtain the most realistic model possible.  By contrast,  John argued for extreme dynamical simplicity, ignoring all the effects of macroscopic motion except the heat flux by thermal convection (yet excluding the momentum and kinetic-energy fluxes), parametrized by a local mixing-length formula, in order to produce a very straightforward model that can be used as a standard benchmark, although in practice that standard has evolved with time, even for John. Here, I shall adopt Model S of \citet{jcdetal1996Sci}  as my reference, because the assumptions upon which it was built have been stated clearly, and because, unlike many other solar models,  it has been produced with sufficient care for reasonably precise, well defined (adiabatic) oscillations to be computed from it. 

I do not go into the details of the early, pre-main-sequence, dynamics. That remains a complicated area of (active) research which continues to benefit from advances in (principally numerical) simulation. Suffice it to say that, because the age of the Sun is very much greater than any of the dynamical timescales, its present structure is hardly likely to have any recall of the precise manner in which the Sun condensed from the interstellar medium, save for the relic chemical composition, and the total mass and angular momentum. It is normally presumed that the chemical composition was either initially uniform, or that mixing during Hayashi contraction homogenized any initial internal inhomogeneity well before the star settled onto the main sequence. However, that assumption is open to question. It is also normally presumed that there was no substantial mass loss nor accretion, although that presumption is also questioned from time to time, with good reason. It is generally believed that, in common with other stars, there has been some loss of angular momentum, particularly in the immediately pre-main-sequence and early main-sequence stages of evolution (and even more before), as is evinced by the decrease with age of the rotational velocities of the photospheres of stars in young clusters  
\citep[e.g.][]{skumanich_clusterrotation1972ApJ...171..565S}. 
However, that process is unlikely to have had a great effect on the state of the Sun today (in other words, it is of little concern that the angular momentum in the very early days might have been rather greater than it is today), and is accordingly rarely discussed. Nevertheless, a word or two about it later in my discussion may not be out of place.   It may have been responsible for a degree of mixing of the products of nuclear reactions in the core, especially in early times. Be that as it may, the essentially static effect of rotation, via the centrifugal force, is very much smaller than the pressure gradient and the force of gravity (as also are Maxwell stresses and the momentum flux due to large-scale convection, except in the very outer, near-surface, layers) so to a first approximation the Sun can well be regarded as being spherically symmetric. Oblateness of the figure resulting from rotation can be treated subsequently as a small perturbation, as I discuss later. So can the effect of mass loss.

\section{Standard main-sequence evolution}
\label{sec:2}
The basic principle behind the evolution is quite straightforward and very robust. Nuclear reactions in the core convert hydrogen into helium, reducing the number of particles -- nuclei, ions  and electrons -- per unit mass of the stellar material, reducing the pressure at a given temperature, permitting the star to contract under gravity and thereby raising the temperature in the innermost regions to maintain hydrostatic balance. The temperature- and density-sensitive nuclear reaction rates are augmented with this contraction by more than the reduction due to the loss of hydrogen fuel, essential to maintain global stability, so the luminosity of the star increases. This basic result is an inevitable consequence of thermonuclear physics and the conditions for hydrostatic balance, and is beyond challenge (provided, of course, that standard nuclear and gravitational physics are accepted). 

The theoretical variation of the luminosity $L$ with time $t$  is given approximately by 
\begin{equation}
L(t)\simeq\frac{L_\odot}{1+\beta(1-t/t_\odot)}\label{2.1}
\end{equation}
\citep{dog1977soiv.conf..451G} with $\beta=L(0)/L_\odot -1\simeq 2/5$, where the origin of time $t$ is at the start of the evolution on the main sequence.  It is robust against minor variations in the assumptions behind the model \citep[e.g.][]{doglakecomo_1988stre.conf...90G,Bahcallpinsonneaultbasu2001ApJ}.  
 A `derivation' of this formula is presented by 
 \citet{doglakecomo_1988stre.conf...90G,1990stromgren}  based on homology scaling, although the constant 2/5 was obtained by adjusting an initial theoretical estimate to account for the  deviation from homologous variation due to nuclear transmutation such as to render equation (1) consistent with numerical stellar-evolution computation,  so the relation should perhaps be regarded  more as a physically motivated interpolation formula. The formula has been generalized to take account of a putative temporal variation of the gravitational constant, $G$, and has also been adapted to accommodate main-sequence mass loss \citep{1990stromgren},  thereby summarizing the numerical computations that have been carried out both before and since. 

One of the reasons for considering a temporal variation in the gravitational constant was to obviate what has been called the faint-Sun problem (some have even called it a paradox). It was posed by \citet{saganmullen_1972Sci...177...52S}, who argued from simple equilibrium energy balance that early in the Sun's main-sequence evolution the Earth would have been completely glaciated had the Sun evolved essentially according to equation (\ref{2.1}), which is counter to geological finding. The solar luminosity is a steeply increasing function of $G$  \citep{teller_G-dependenceofLsun_1948PhRv...73..801T}: $L \propto G^{7.8}$ \citep{gamow_G(t)1967PNAS...57..187G}.  Therefore adopting an appropriate small decline of $G$ with time can almost annul the rise in $L$ described by equation (\ref{2.1}): if, for example, $G(t')/G(t_{\rm u})=(t'/t_{\rm u})^{-q}$ (where $t'$ is time measured from the Big Bang and $t_{\rm u}$ is the age of the Universe), then the irradiance on Earth would 
have been very nearly constant\footnote{having deviated from its initial value by no more than 0.5 per cent.}   if $-G/{\dot G}|_{t'=t_{\rm u}}\simeq 1.2\times10^{11}\,{\rm y}$,  almost irrespective of the value of $q$ \citep[e.g.][]{doglakecomo_1988stre.conf...90G,1990stromgren}. However, this is not strong evidence in favour of there being a temporal variation in $G$, because it is unlikely that the climatological equilibrium energy-balance assumption is correct.  Apparently more sophisticated meteorological arguments have even been perceived to exacerbate the issue. For example, in the summer of 
1973, motivated by a prediction that the Sun's luminosity might have declined suddenly by a few per cent  some million or so years ago \citep{dilkegough1972Natur.240..262D}, Tzvi Gal-Chen and I (unpublished) engineered a 5-per-cent reduction in solar irradiance in the Global Circulation Model (GCM) of the Earth's 
atmosphere   \citep{Kasahara_Washington_GCM_1971JAtS...28..657K}  at the US National Center for  Atmospheric Research. As expected, the Earth became completely glaciated; but what was not expected is that when the irradiance was subsequently restored to its present value, the ice that had just been produced 
would not melt. That result is evidently contrary to observation, because the Earth is not completely glaciated today. The moral, for me at least, was to confirm that one must always beware of the products of complicated 
computer programmes, particularly when they are state of the art yet depend, as does the GCM,  on rather primitive physics. The many physical processes that the GCM had to account for were ill understood -- as indeed many are still -- and their influence on the atmospheric dynamics had to be parametrized; the controlling parameters whose values were not previously known were measured, where possible, and then 
held constant in the model. That was the most obvious flaw. We know that the stability and evolution of any dynamical system can depend quite sensitively, and sometimes critically, on the nature of the constraints that are imposed; evidently a meteorological model, however sophisticated, with processes whose controlling 
parameters do not change appreciably on a timescale of days or weeks could not possibly be expected to apply to the climate on timescales of millions of years. Indeed, Dilke and Gough had presumed that the response of the Earth's climate to irradiance variation was quite different on a timescale of $10^6$y  from that on a timescale of $10^8$y. The processes responsible for such timescale dependence have been studied 
subsequently by, amongst many others,  \citet{margulislovelock1974Icar...21..471M}, \citet{margulislovelock1974OrLi....5...93L} and \citet{Gaia_review1982P&SS...30..795L}, who likened the long-term stability of the climate to the work of Gaia, the ancient Greek goddess of Earth. The cause of the suspected sudden relatively recent decline in $L$ was itself a dynamical process, considered to have resulted from a nonlinear instability of g modes in the Sun's core, the theory of whose development also depending critically on what is held constant, a matter to which I shall return later.  

\begin{figure}
\centering
\includegraphics[width=0.85\linewidth]{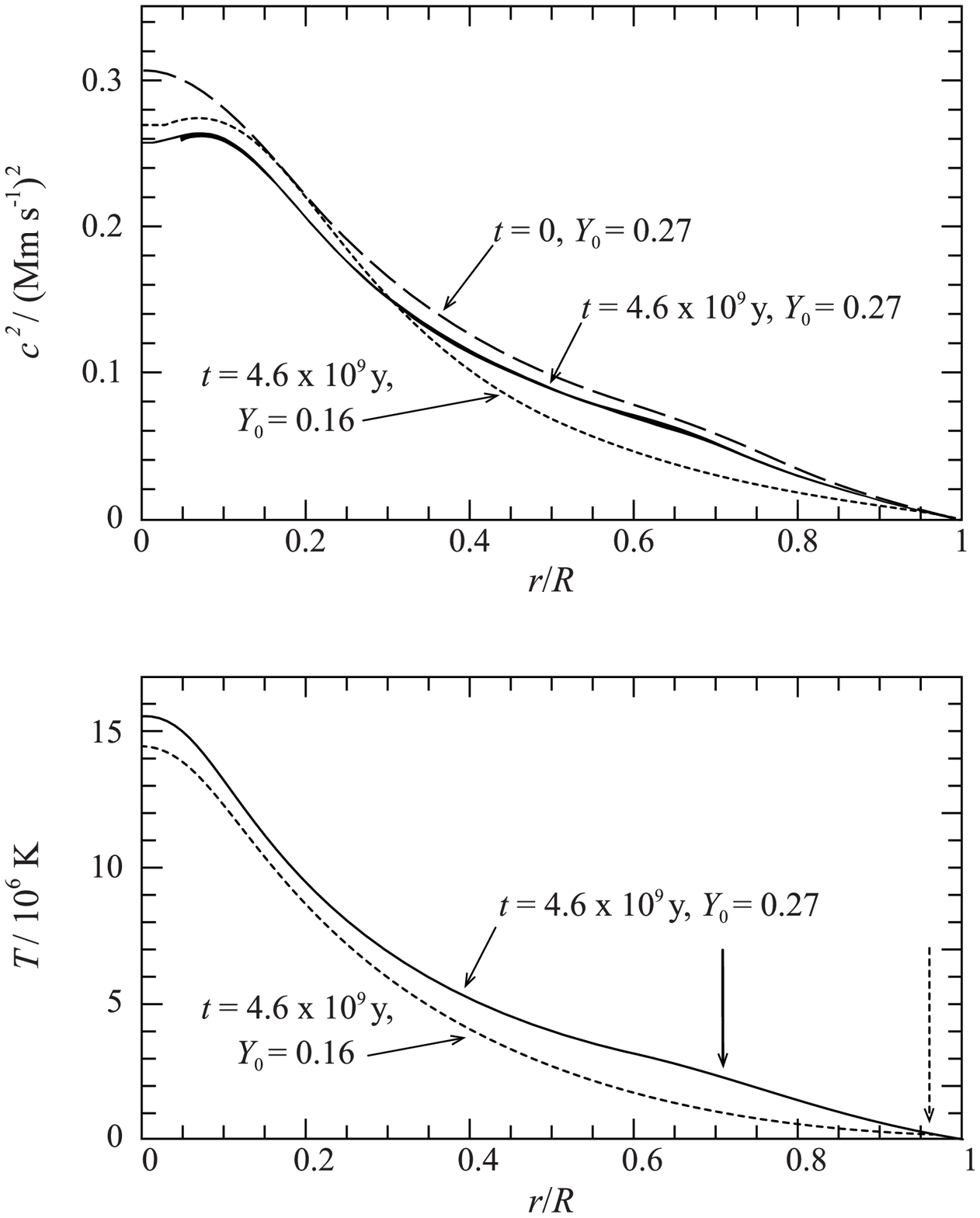}
\caption{Square of the sound speed,  $c^2$, and temperature, $T$, in two solar models computed with a local mixing-length formalism and having different initial heavy-element  abundances $Z_0$: one (continuous curves) is the standard 
Models S of \citet{jcdetal1996Sci}, having $Z_0 = 0.020$ and a corresponding initial helium abundance $Y_0 = 0.27$; the other (short-dashed curves) has 
$Z_0 = 0.001$, $Y_0 = 0.16$ and has been continuously contaminated with heavy elements at the the surface at such a constant rate as to have a surface abundance $Z_{\rm s}=0.020$ today \citep{dirtymodels1979A&A....73..121C, dirtymodelscorrigendum1979A&A....73..121C}. The 
depression in $c^2$ (relative to $T$) in the central regions of the cores of the evolved models is a result of the augmentation of the mean molecular mass in the core produced by the transmutation of hydrogen into helium. Included, for comparison, as a long-dashed curve, is the square of the sound speed of the higher-$Y$ model at zero main-sequence age, which, of course,  has no such depression. The vertical arrows (with line styles corresponding to the models to which they refer) mark the bases of the convection zones of the two present-day models, where the second derivatives of $c^2$ and $T$ are discontinuous. The square of the sound speed in the Sun, inferred seismologically, 
is drawn also as a continuous curve, not quite reaching the centre $r/R=0$ of the star; it is barely distinguishable from $c^2$ in the higher-$Y$ model (after \citet{dog1999NuPhS..77...81G}).
} 
\label{fig1}
\end{figure}

The computation of the structure and evolution of (standard) solar models is straightforward once the equation of state, the nuclear reaction rates and the opacity of the stellar material have been specified. There is also a matter of determining the mean stratification of the convection zone: to that end some form of mixing-length formalism is normally used to specify the energy (usually just heat) flux and, sometimes, the Reynolds stress\footnote{yet hardly ever the flux of kinetic energy}. For most purposes the deficiencies in such a procedure are of little concern to the overall structure of the star, because throughout almost all of the convection zone the stratification is adiabatic and the Reynolds stress is negligible; it is only in the outer boundary layer where the details of the formalism matter.   In practice the mixing-length formalism contains at least one explicitly adjustable constant which can be calibrated to determine the value of the adiabatic constant deep in the convection zone that is required to reproduce the observed radius of the Sun; calibrated in that way, the outcome is a function of chemical composition, and it formally determines the depth of the convection zone.  Of course, that is not to say that in reality the dynamics of the upper convective boundary layer depends explicitly on the chemical composition in the manner determined by the solar calibration. 

The chemical composition is normally specified by two of three parameters: the relative abundances $X$, $Y$, $Z$ by mass of H, $^4{\rm He}$ and of all other elements combined\footnote{usually referred to as `heavy elements', even though they include $^3{\rm He}$}; they satisfy $X+Y+Z=1$. (The relative abundances of the elements incorporated into $Z$ are provided principally, but not completely,  by spectroscopic analyses of the Sun's atmosphere \citep{Asplundetal2009ARA&A,caffauetal2009A&A}.) Stellar evolution theory provides a further relation between them from the requirement that the current luminosity $L (t_\odot)$ agrees with the value measured (assuming the age $t_\odot$ to be known). That leaves a single infinity of apparently acceptable models, which can be labelled by any one of $X$, $Y$ and $Z$. I record that along the sequence of these models, both $Y$ and the depth $d_{\rm c}$ of the convection zone increase monotonically with $Z$. 

Before helioseismology, there was just one measured quantity associated with the Sun that in principle could be used to select the most appropriate model, namely the neutrino ÔluminosityÕ $L_\nu$ (usually expressed as an associated flux $F_\nu$ at 1AU computed under the assumption that neutrinos do not decay, which itself is expressed as a capture rate by whatever neutrino detector is being considered). Theoretically, $L_\nu$ is a monotonic increasing function of $Z$. Therefore a unique model can thereby be chosen. However, that model left many people uncomfortable, and was deemed unacceptable by most astronomers and astrophysicists because the value of $Y$ in that model was much lower than that observed in open stellar clusters containing apparently Sun-like stars; moreover, it was lower than the amount of helium created in generally accepted models of Big-Bang nucleosynthesis. Astronomical opinion prevailed, and the discrepancy posed what was accordingly named the solar neutrino problem\footnote{rather than, for example, the solar abundance problem}. 

Perhaps the first genuinely seismological inference to be drawn about the structure of the Sun was from a crude calibration of the depth  $d_{\rm c}$ of the convection zone 
\citep{DOG1977Nice}. It implied a value of $Y$ even greater than those preferred at the time, exacerbating the neutrino problem. The inference was soon confirmed with more detailed, numerical, 
calculations by \citet{rhodesulrichsimon1977ApJ...218..901R}, 
and later by 
\citet{nice5minpaper1980LNP...125..307B} and \citet{lubowrhodesulrich1980LNP...125..300L},  who addressed the sensitivity of the eigenfrequencies to uncertainties in the background model envelope.
Subsequently, a helioseismological 
inference of the sound speed throughout most of the solar interior \citep{speedofsoundJCDetal1985Natur}  
not only provided a consistent direct measurement of $d_{\rm c}$ but also ruled out the 
$F_\nu$-reproducing low-$Y$ model. A more recent, and cleaner, demonstration of that inference is reproduced in Figure 1. It convincingly demonstrated that the 
resolution of the solar neutrino problem must lie in nuclear or particle physics, and not in stellar physics, although that conclusion was not accepted immediately by most of the heliophysics community who did not yet appreciate the power of helioseismology  \citep[e.g.][]{bahcallulrich1988RvMP.60.297B}.   I shall say no more about that here because it is only indirectly associated with interior dynamics (but see \citet{CummHaxton1996PhRvL..77.4286C}.)

\section{Angular velocity and the solar oblateness}
\label{sec:3}
The most obvious indicator of global solar dynamics is the angular velocity $\Omega(r,\theta,t)$  (I adopt spherical polar coordinates  $(r,\theta,\phi)$). A seismological determination of $\Omega$ in the Sun's envelope by \citet{schouetalrotation1998ApJ} is illustrated in Figure 4, and a schematic representation by \citet{chaplinetalrotation1999MNRAS.308..405C}  extending to the centre is plotted in Figure 2. Roughly speaking, the angular velocity is independent of radius and increases with colatitude $\theta$ in the convective envelope $(r>r_{\rm e})$, and it is approximately uniform in the radiative envelope beneath, except possibly in the energy-generating core $(r<r_{\rm c})$ where there is (slight) evidence\footnote{which is commonly doubted}  that the spherically averaged $\Omega$ is lower than $\Omega$ in the radiative envelope \citep{elsworthetalrotation1995Natur.376..669E}. The transition at the base of the convection zone, called the tachocline  (Spiegel and Zahn, 1992; see also \citet{eastachycline1972NASSP}), is too abrupt to be properly resolved seismologically, as is the transition, if there is one,  at the edge of the core. 

It is common practice to expand the latitudinal dependence of $\Omega$ in even powers of $\mu={\rm cos}\theta$:
\begin{equation}
\Omega(r,\theta,t)=\sum_{k \ge 0} \Omega_k(r,t)\,\mu^{2k}\,, \label{3.1}
\end{equation}
where, on the whole, $\Omega_k$ varies weakly with $r$ and $t$, except in the tachocline\footnote{Often, for computational convenience, $\Omega$ is expanded in orthogonal polynomials, such as Clebsch-Gordon coefficients  \citep{ritzwollerlavely1991ApJ...369..557R}}.   It is interesting and, so far as I am aware, unexplained, that for most values of $r$ the magnitude of $\Omega_k$ is particularly small for all $k>2$. The coefficients of the terms of lower  degree  are observed to vary somewhat with the sunspot cycle, but that variation has been convincingly detected only in the convection zone, so I refrain from discussing it in any detail here. There is also a variation with a characteristic period of about 2 years \citep{BiSON_QBO_2012MNRAS.420.1405B}, which has been potentially misleadingly called the quasi-biennial oscillation --  it can hardly be compared with the terrestrial disturbance with the same name, having the acronym QBO, because the latter is driven by differential gravity-wave dissipation, a process which I explain in \S10, and which can hardly be sustained in the convection zone of the Sun, where gravity waves cannot propagate.

\begin{figure}
\centering
\includegraphics[width=0.85\linewidth]{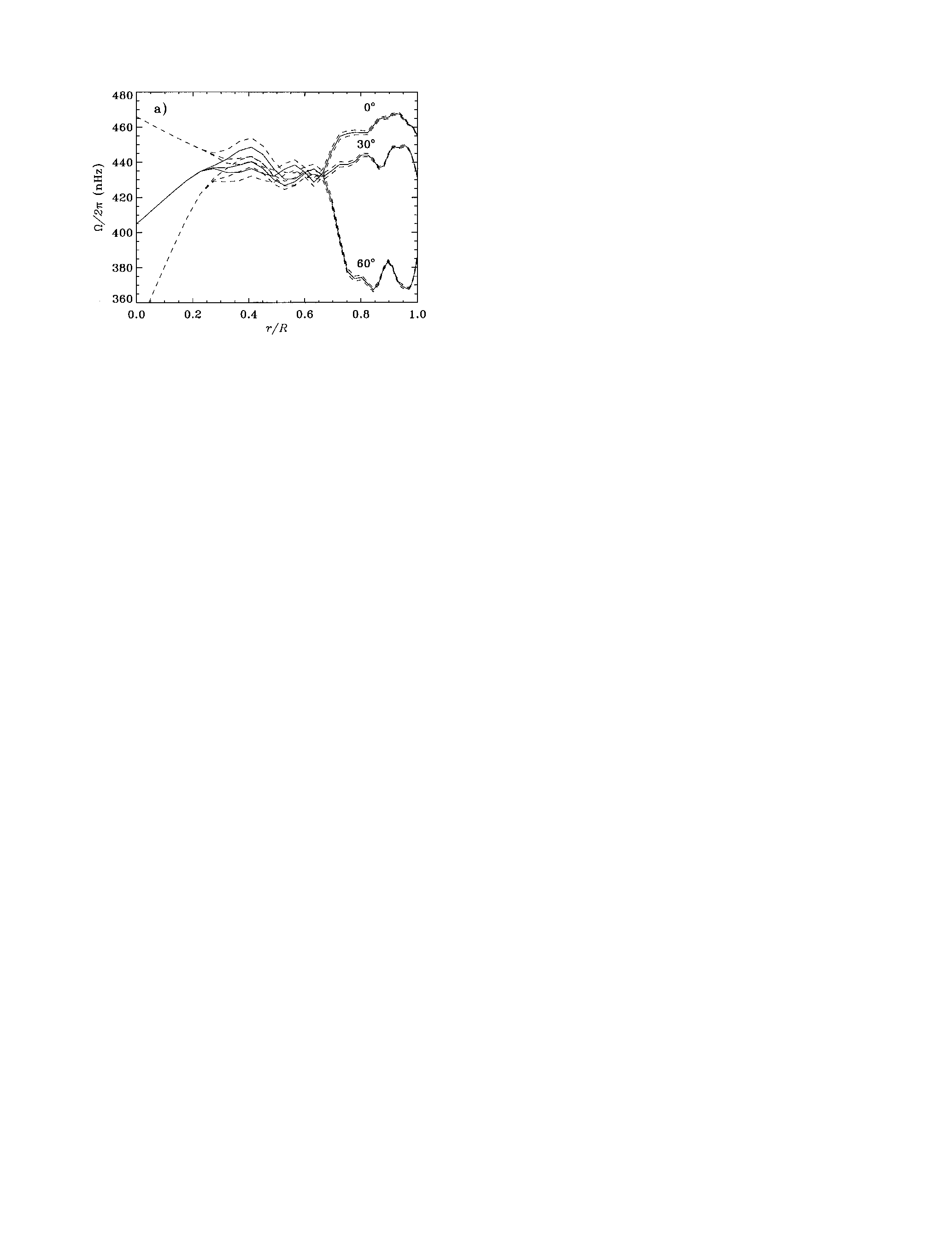}
\caption{Angular velocity through the Sun, plotted against radius at three different latitudes. The continuous curves are the expectations (averages), and are flanked by dashed curves with added or subtracted standard errors propagated from the estimated errors in the data. The tachocline is not well resolved; nevertheless its base appears to be spherical, as one would expect,  and penetration of the shear into the convection zone 
preferentially at high latitudes, rendering the overall shear layer prolate, is discernible. The angular velocity in the radiative envelope (outside the energy-generating core) is essentially uniform, within the uncertainty, and there is a hint that the  core is rotating more slowly \citep[from][]{chaplinetalrotation1999MNRAS.308..405C}.  The confluence at  $r/R \simeq 0.25$ of the values of $\Omega$ at the three different latitudes and the almost linear dependence beneath are both consequences  of the regularization, which, because the seismic data are too scant to resolve the angular velocity well, dominates the inversion procedure. 
All that can be inferred is that an equatorially biassed latitudinal average of the angular velocity in the core appears likely to be lower than the angular velocity of the radiative envelope 
\citep[cf.][]{elsworthetalrotation1995Natur.376..669E}; therefore the apparent uniformity of $\Omega$ with respect to latitude in the core  cannot be regarded as evidence for an absence of latitudinal variation.
} 
\label{fig2}
\end{figure}

The first conclusion to be drawn from the earliest determination of $\Omega$ was an estimate of the quadrupole moment $J_2$ of the external gravitational field \citep{Naturerotation1984Natur}. It is induced by the centrifugal force acting on the stellar material, causing the mass distribution to be oblate. Therefore $J_2$ can be represented as an appropriately weighted average of $\Omega^2$ over the volume of the star \citep{DOGrotation1981MNRAS.196..731G, pijpers1998MNRAS.297L..76P}, the weight function $F$  being determined by linear perturbation about the nonrotating state.   The  radial dependence of  the component of 
$F$ pertaining to the spherically averaged $\Omega$ is plotted in Figure 3. Knowledge of $J_2$ is essential for testing theories of gravity from measurements of planetary precession and spacecraft orbits, for non-Newtonian theories produce a perturbation to Newtonian orbits that at present are observationally indistinguishable from that which is produced by a 
deviation from spherical symmetry induced by an appropriate internal centrifugal force.

Early attempts to measure $J_2$ were made by 
\citet{auwers_oblateness_1891AN....128..361A}, \citet{AmbronnSchur_oblateness1905} and 
\citet{Poor_oblatenessI_1905ApJ....22..103P, Poor_oblatenessII_1905ApJ....22..305P} from direct observations of the presumedly centrifugally induced visual oblateness, defined as   
$\Delta_{\rm v}=(R_{\rm e}-R_{\rm p})/R$, where $R_{\rm e}$, $R_{\rm p}$ and $R$ are respectively the 
equatorial and polar radii and their average\footnote{The oblateness $\Delta_{\rm v}$ depends also 
on $J_4$ and the higher moments, but 
the additional contributions appear to be less than the observational uncertainty, so for clarity I do not take them explicitly into account here.}; in retrospect those  observations were inconclusive.  More modern 
measurements  by \citet{DickeGold1967PhRvL, dickegoldenberg1974ApJS} suggested that  $\Delta_{\rm v}$ is about 5 times greater than what one would expect from a distortion by a uniform rotation of the Sun's interior,  consistent with the \citet{BransDicke1961}  theory of gravity, appropriately calibrated.     However, that was challenged by  \citet{hill-stebbins_oblateness1975ApJ...200..471H}, whose measured value was consistent with uniform rotation.   
These and subsequent measurements have been reviewed by, for example, \citet{rozelot-etal_oblateness2011JASTP..73..241D}.  
The shape measurements were very difficult, partly because $\Delta_{\rm v}$ is very small, of the order of the centrifugal parameter $\Lambda=R^3\Omega^2/GM\simeq 2\times10^{-5}$, where $M$ is the mass of the Sun, and partly because we now know that the oblateness $\Delta_{\rm v}$ of the visible solar disc is dominated not by the oblateness $\Delta_\Phi$ of the gravitational field caused by the action of the centrifugal force on the dense interior material, but principally by the oblateness $\Delta_\Omega$ caused by the direct effect of the centrifugal force on the diffuse visible surface layers. Therefore $\Delta_\Phi$  would need to be determined as the relatively small difference between $\Delta_{\rm v}$ and $\Delta_\Omega$, which is an intrinsically unreliable procedure. The difficulties in the entire measurement procedure have been  compounded by the fact that the relation between the oblateness of the observed radiative intensity and the oblateness of, say, the surfaces of constant pressure in the photosphere is contaminated with brightness variations due to sunspots and magnetically generated excess emission, such as faculae. Accounting for these is an uncertain process, as is evident from even the most recent investigations \citep{solarcycleradiuschanges2007ApJ.658L.135L, fivianetal2008, kuhnetalsolarshapeScience2012}. 

\begin{figure}
\centering
\includegraphics[width=0.85\linewidth]{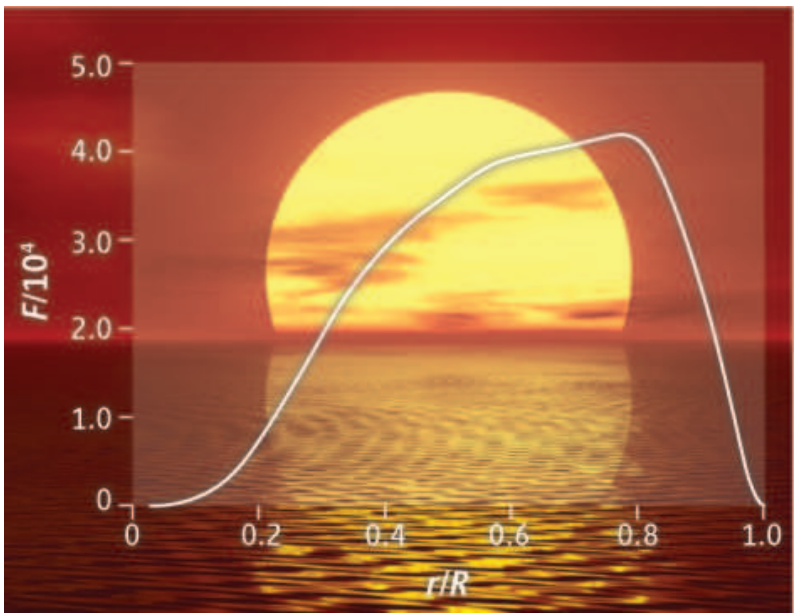}
\caption{Superposed on the setting Sun is a plot of the weighting function $F(x)$ in the approximate formula $J_2  \simeq \int F \overline{\Omega}^2{\rm d}x$, where $x=r/R$ and  $ \overline{\Omega}$ is the spherical average of the (latitudinally mildly varying) internal angular velocity of the Sun, in units of ${\rm s}^{-1}$ \citep[after][]{dogscienceprospects2012}.    The function $F$ is small near the centre of the Sun, where the ratio of centrifugal force to angular velocity is low, and near the surface where the low density, however oblate, makes only a minor imprint on the gravitational potential. 
} 
\label{fig3}
\end{figure}

The advent of helioseismology changed the situation dramatically, because the seismic signature pertinent to determining $J_2$ can be measured much more accurately: that signature is the rotational frequency splitting, whose magnitude is of order $2m\Omega$, where $m$ is the magnitude of the azimuthal order of a seismic mode; modern measurement error of mean dipole-mode splitting, for example, is approximately $1\%$ \citep{Howe2009}, which offers some idea of the precision to which $\Omega$ can be inferred. Furthermore, the dynamical relation between $\Omega$, which is inferred directly from the frequency splitting, and $J_2$ is not influenced significantly by non-dynamical variables such as radiative intensity. From even the first primitive measurement of the interior angular velocity \citep{Naturerotation1984Natur} it was evident that the prediction of the non-Newtonian component of the precession of the perihelion of the orbit of Mercury is consistent with General Relativity.  Subsequent measurements, such as those reported by \citet{schouetalrotation1998ApJ}, tightened that conclusion, and will in future provide more stringent tests of theories of gravitation once finer orbital measurements become available. 

It should be noted that the only unambiguous seismic signature of rotation comes from inertial (such as Coriolis) effects on the nonaxisymmetric  ($m>0$) modes, which are tiny in the evanescent region of those modes:  namely, near the axis and, doubly, near the centre (because the degree $l$ cannot be less than $m$), which explains the white region in Figure 4.  That region coincides, accidently, with one of the regions in which  the oblateness kernel (whose spherical-average component is illustrated in Figure 3) is small, because the centrifugal force is small.  Therefore the principal uncertainty in the inference of $\Omega$  contributes little to the uncertainty in the value of $\Delta_\Phi$ that is derived.

\begin{figure}
\centering
\includegraphics[width=0.85\linewidth]{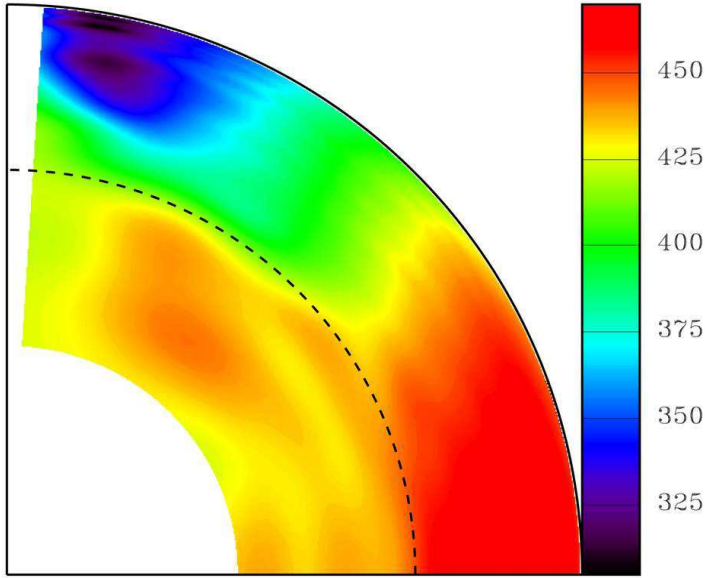}
\caption{Colour rendering of the rotation rate of a meridional quadrant of the Sun. The scale is in nHz. The continuous quarter circle is located in the photosphere, the dashed quarter circle at the base of the convection zone. The white area denotes the region in which the rotational splitting data provide only scant indication of the angular velocity (from the MDI Image Gallery: http://soi.stanford.edu/results/Triana/modslice.ps, constructed from the work of \citet{schouetalrotation1998ApJ}). 
} 
\label{fig4}
\end{figure}

\section{A dynamical issue raised by recent oblateness measurements}
\label{sec:4}
Taken at face value the most recent direct measurements raise an interesting dynamical question. After doing their best to remove facular contamination by rejecting excessively bright regions at the solar limb, which are equatorially concentrated and therefore contribute positively to the oblateness of the intensity distribution, \citet{kuhnetalsolarshapeScience2012} found a residual visual oblateness that is actually less than $\Delta_\Omega$ by about $8\times10^{-7}$. The observations by Fivian et al. (2008) led to a similar, although less extreme, result\footnote{In their paper, \citet{fivianetal2008} actually claim a value that exceeds $\Delta_\Omega$ by about $2\times 10^{-7}$, which they offer as an estimate of $\Delta_\Phi$, but their result was obtained with  Dicke's (1970) outdated underestimate of $\Delta_\Omega$.}. So is the Sun gravitationally prolate? The only way that that could be is probably for the radiative interior to be constricted by a predominantly toroidal belt of magnetic field (with an associated poloidal component, for stability) encircling the rotation axis, of intensity that would need to be $10^4$T or more. Is that possible? What are the alternatives?

Of course there is always the possibility that the observations were inadequately analysed. Kuhn et al. explicitly rejected bright facular regions. Did they take adequate care to reject darkened, presumably  penumbral regions whose inappropriate inclusion could have rendered the radiative intensity more prolate? Fivian et al. accounted for faculae by using a proxy indicator for rejecting data and extrapolating $\Delta_{\rm v}$ to 100\% rejection; there is a fear in some quarters that the resulting statistics are inadequate for rendering the outcome reliable. A na\"ive theorist insufficiently familiar with the details of the observations might simply take the difference between the two inferred values of the uncontaminated visual oblateness, about $8\times10^{-7}$, as a plausible estimate of the uncertainty, and, noticing that it is nearly three times the value of $\Delta_\Phi$ obtained from helioseismology, set the matter aside. 

It is nonetheless interesting to entertain the idea that there is a prolate contribution to the distortion of the surface. As \citet{dicke1970ApJ} has emphasized, that must necessarily occur in the seen photospheric layers. A solar physicist's immediate reaction is likely to be to invoke a superficial magnetic field. The photospheric mean (spatially smoothed) field appears to be principally poloidal and dipolar at most times, and of insufficient intensity to maintain the pole-equator photospheric radius difference of the order of the $10^{-6}R$ that would be  required to explain the findings of \citet{kuhnetalsolarshapeScience2012}.  Moreover, it would enhance, not reduce, the visual oblateness.  On the other hand, a preferential suppression of polar convection by only  a mere $10^{-3}$ per cent, which is arguably more plausible, would be sufficient to induce the required reduction in oblateness if the local static adjustment of the star in only the vertical direction were to matter.  What would be the dynamical consequence? The poles would be elevated relative to a surface that is everywhere perpendicular to the combined gravitational and centrifugal forces, and if convective Reynolds stresses acted like a scalar viscosity, which is not uncommonly assumed yet which is almost certainly incorrect, photospheric matter would flow downhill towards the equator, which is contrary to the direction observed. The dynamical problem is evidently not straightforward. Yet it is an interesting and evidently important problem, for, irrespective of the oblateness issue, a cogent explanation of the poleward meridional flow in the outermost layers of the convection zone is lacking \citep[e.g.][]{ToomreThompsonISSI2015}.

\section{Spin-down}
\label{sec:5}
The solar wind is rotationally coupled to the Sun via a large-scale magnetic field. It is caused to rotate roughly at the solar rate out to about 5 solar radii, thereby removing angular momentum from the Sun; beyond, angular momentum in the wind is more-or-less conserved. Thus the rotation of the outer layers of the Sun must be decelerating, in common with inferences for other stars drawn from observations of rotational spectral-line broadening in open clusters of difference ages \citep{skumanich_clusterrotation1972ApJ...171..565S}. An interesting dynamical question that arises from this process is how far into the Sun this deceleration penetrates. Of course we now know the answer from seismological measurement, but the dynamical issues, to some extent, remain. 

The matter was debated in the late 1960s after \citep{dicke1964Natur} had tried to maintain that the Sun's core is rotating so rapidly (with a period of a little over a day) as to induce a gravitational oblateness of magnitude sufficient to sever the agreement between the observed residual precession of the orbit of Mercury and the prediction of General Relativity. Dicke pointed out that the global viscous diffusion time $\tau_{\rm v} = R^2/\nu$ (where $\nu$ is  a mean kinematic viscosity) exceeds even the age of the Universe, and that therefore the Sun's core is rotating essentially at its pre-main-sequence rate. He argued that
the shear could be stable to the Richardson criterion, so shear turbulence would not add to the viscous stress, and that Maxwell stresses could be insignificant too.  \citet{howardmoorespiegel67} and \citet{brethertonspiegel1968ApJ} pointed out a fundamental flaw in Dicke's claim, using simple dynamical models to demonstrate that meridional advection, and not viscous stress, is likely to be the dominant agent transporting angular momentum through the body of the Sun, in a process called spin-down \citep{greenspanhoward_spindown1963JFM....17..385G}, as in a stirred cup of tea.

Interestingly, the spin-down process had been discussed qualitatively long before by \citet{einsteinmeanders} as an explanation of the meanders of rivers (unwittingly defending his theory of General Relativity), and quantitatively by \citet{bondilyttleton_spindown1948PCPS...44..345B}  in a discussion of the deceleration of the Earth's core.
{\begin{figure}
\centering
\includegraphics[width=0.85\linewidth]{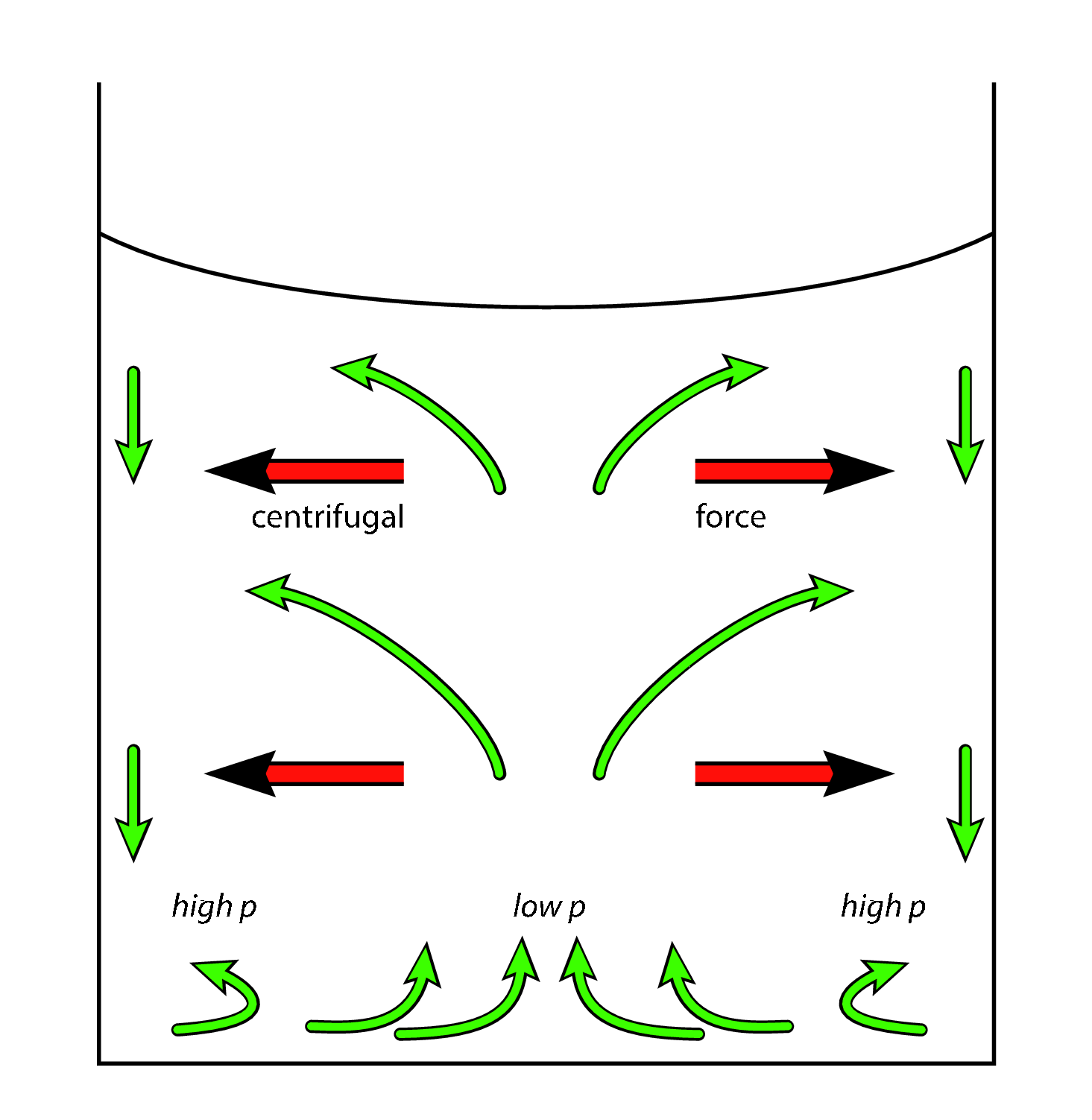}
\caption{Spin-down in a stirred cup of tea. The free surface of the fluid is denoted by the curved black line, the cylindrical vessel containing the fluid by the vertical and horizontal lines. The thicker arrows (red) with black heads indicate the centrifugal force acting on the body of the rotating fluid; that force is essentially absent from the Ekman boundary layer near the base of the container, which hardly rotates. The thinner (green) arrows indicate the direction of the induced meridional flow, which, in the main body of the fluid, is essentially inviscid and therefore angular-momentum conserving. Angular momentum is lost from the fluid principally by viscous transfer in the thin Ekman boundary layer.  Such a boundary layer is not essential to the global mechanism of spin-down, however: as \citet{brethertonspiegel1968ApJ} have argued, penetration of the meridional flow into the Sun's convection zone could provide an (even more) effective mechanism for divesting angular momentum.  That process can be modelled  on Earth by spreading a layer of heavy beads over the bottom of the teacup.
} 
\label{fig5}
\end{figure}
I describe the broad principles briefly here, because, as I shall mention later, the general conclusion may find astrophysical application elsewhere. 

Consider a stationary cylindrical vessel of radius $R$ containing water initially rotating (approximately) uniformly  with angular velocity $\Omega$ about its vertical axis, a situation discussed, for example, by \citet{greenspanrotatingfluids1969}. The situation is illustrated in Figure 5. The fatter (red), horizontal, double arrows represent the centrifugal force, which is balanced by a pressure gradient directed away from the axis. The excess pressure $\Delta p$ near the outside wall of the container supports a greater head of water, and therefore the 
upper, free, surface of the water is concave upwards. Surfaces of constant pressure are not normal to gravity, a situation known as baroclinicity.  Viscous stresses slow the rotation of the water near where it touches the container, particularly at the bottom, thereby reducing the centrifugal force.  That leaves an unbalanced 
component of the pressure gradient which drives a spiralling inward flow with characteristic speed $u$,  indicated in Figure 5 by the thinner (green) arrows, in a thin boundary layer (of characteristic thickness $\delta \ll R$), now called an Ekman layer 
\citep[after][]{ekmanlayer1905}, near the bottom of the container.   It is there that angular momentum is removed from the water\footnote{Viscous stress operates also on the side walls, but there the boundary layer is not as thin as  that at the bottom of the container, and removes negligible angular momentum (just as Bondi and Lyttleton had found, in the case of spin-down in a sphere, that negligible angular momentum is removed near the equator). The bottom boundary layer is thinner as a result of the vertical `rigidity' imparted on the fluid immediately above by the vortex stretching  (which is intimately related to a tendency towards local angular-momentum conservation) 
produced by the shear, and which is also responsible for the better-known Taylor-Proudman theorem for steady incompressible inviscid flow.}. In the boundary layer the Coriolis force (in a frame rotating with the bulk of the fluid) is balanced by the viscous stress in the shear:  $\Omega u \simeq \nu \delta^{-2} u$.  The radial pressure gradient,  of order $\Delta p/R$, is also balanced by viscous stress, which determines the boundary-layer velocity: $\Delta p/R \simeq R \Omega^2 \simeq \nu u / \delta^2$.  The return flow -- upwards and outwards in the bulk of the fluid where viscous stresses are negligible, with characteristic velocity $w \simeq u\delta /R$ -- slows the rotation in the bulk of the fluid mainly by angular-momentum conservation, on the timescale $\tau \simeq R/w$. 
This  spin-down (equilibration) time is thus $\tau \simeq R^2/\delta u \simeq R/\sqrt{\nu \Omega}$;  it is the geometric mean of the global viscous diffusion time $\tau_{\rm v} = R^2/\nu$ and the characteristic dynamical time $\Omega^{-1}$, which is much shorter than the viscous time.   More accurate formulae for $\delta$ and $\tau$ can be obtained by a local analysis of the boundary layer, which relates the vertical shear in the angular velocity to the upflow velocity out of the Ekman layer \citep{greenspanrotatingfluids1969} consequent on what is now known as gyroscopic pumping. 

An important observation of the process is that the flow near the bottom of the container is inwards, towards the central axis. That is why loose tealeaves in a stirred cup of tea migrate to the middle of the cup, and silt at the bottom of a slowly flowing, meandering,  river is transported from the outer to the inner bank of a bend, accentuating the meander \citep{einsteinmeanders}. It is evident from the balance of forces that if instead the container were caused to rotate more rapidly than the fluid, a flow with the same geometry would be induced, but with its direction reversed. 

The principal importance of this process for spin-down is not in the details of the Ekman layer, but simply in the fact that the layer 
extracts angular momentum from the fluid, leaving an unbalanced pressure gradient within the layer to drive a large-scale flow towards (or away from, were the teacup to be rotating faster than the tea) the axis.  I have included the account of the boundary layer here, however,  simply to complete my discussion of some interesting physics.

Returning now to the discussion of the Sun, one must first beware, quite generally, of attributing the properties of oversimplified models to complicated situations without careful consideration of the implications of those simplifications, especially when dynamics is involved. Good models might exhibit some of the physical processes in operation, but it must always be appreciated that they may be no more than merely illustrative. 
It is therefore not uncommon for arguments based on the understanding gleaned from them to be challenged, 
in disbelief of the extension of the domain of  applicability  necessarily required for addressing the matter in 
hand.  

For example, \citet{dickespindown1967ApJ} pointed out that the Sun is no cup of tea, and that the base of the convection zone does not exert a stress in the manner of the base (or top) of a rigid container to produce a diffusive Ekman layer.  \citet{brethertonspiegel1968ApJ} countered by explaining that an Ekman layer is not essential: 
the pertinent agent is the induced meridional flow which advects angular momentum essentially inviscidly throughout the body of the fluid; that flow can penetrate into the convection zone where it can divest its angular momentum even more efficiently than in a viscous boundary layer. They illustrated the process by modelling 
the convection zone as a rigid porous medium.\footnote{and subsequently demonstrated it in the laboratory (unpublished) in a rotating beaker of water containing several layers of glass beads.}  They modelled the radiative zone as an incompressible fluid, since the details of how exactly angular momentum is transported in the essentially inviscid radiative interior, be it compressible or not,  is of relatively minor importance.

Dicke also tried arguing that the stable stratification of the radiative interior performs the  crucial role of  inhibiting  large-scale, angular-momentum-advecting flow -- as indeed had  \citet{howardmoorespiegel67} recognized already -- confining it to a shallow \citet{holton1965a_JAtS...22..402H,holton1965bJAtS...22..535H} layer immediately beneath the convection zone, thereby insulating the core from the deceleration of the surface (a conclusion which is valid only on timescales shorter than the shortest diffusion -- here thermal -- timescale);  \citet{McD&Dicke_spindown_1967Sci...158.1562M} illustrated the process experimentally, 
concluding that the stratification of the Sun should preclude core spin-down.  \citet{clark2thomas_solar_spin-down_1969Sci...164..290C} and \citet{modisette_novotny_solar_spin-down1969Sci...166..872M} concurred.  Apparently oblivious of the arguments that had been presented by  \citet{howardmoorespiegel67}, Dicke had failed to recognize  that actually the impediment  is thermally moderated --  
negative buoyancy generated by vertical adiabatic motion can be annulled by thermal diffusion -- so one must estimate the buoyancy annihilation time in order to assess the validity of the simplified model.   \citet{roxburgh_oblateness_1964Icar....3...92R} had already pointed out the importance of thermal diffusion,  positing that the diffusively moderated spin-down time is the (thermally driven) Eddington-Sweet timescale, 
which is greater than the age of the Sun.  But Howard, Moore and Spiegel considered, more realistically, the thermal control of the (dynamically driven) flow, estimating the spin-down time to be at most of order only $10^9$yr}.  To fluid dynamicists at the time of the 1960s debate, the balance of the evidence seemed to favour substantial global spindown,\footnote{perhaps even in the face of the steep gradient of molecular weight \citep{HEHEAS_mu_barrier1977ApJ...213..157H} which had previously been regarded as isolating the angular momentum of the energy-generating core \citep{mestel_mu_barrier1953MNRAS.113..716M,mestel_mu_barrier1957ApJ...126..550M}.} 
but the case had not yet been proven.   Subsequent analysis by \citet{easjpztach1992A&A} of a similar, though not identical, situation relating to the seismologically observed near-uniform rotation of the radiative interior in the face of a differentially rotating convection zone, to which I turn my attention in the next section, has added substantial dynamical support to the evidence.  And, of course, the seismological findings themselves negate the view that purported weakness of spin-down has enabled the Sun to have sustained a high gravitational oblateness, although they do not tell us how. 

The situation is changed dramatically once it is recognized that the radiative envelope might be pervaded by a large-scale magnetic field.  Mestel and Weiss (1987; see also \citet{mestel_mu_barrier1953MNRAS.113..716M,mestelbook1999stma.book.....M} and \citet{Cowling_book1976magn.book.....C}) 
have argued that a quite modest poloidal field (about $5\times 10^{-6}$T) is sufficient to connect the core to the convection zone in the lifetime of the Sun, and have advanced arguments to suggest that the actual field  could be of order 10$^{-2}$T. That field is substantially weaker than the only {\it ab initio}, albeit poor, estimate  -- of order 30T -- \citep{dog_PhilTrans_1990RSPTA.330..627G} that I know.  The magnetic spin-down process was subsequently studied numerically by \citet{CharbonneauMacGregor1992ApJ...397L..63C,CharbonneauMacGregor1993ApJ...403L..87C,CharbonneauMacGregor1993ApJ...417..762C}  with an idealized model having a rigidly rotating convection zone. That study confirmed that magnetic spin-down is efficient, but the model was too simplistic to explain why the radiative interior rotates uniformly, an issue to which I now turn.

\section{The steady laminar tachocline}
\label{sec:6}
The angular velocity of the convection zone varies with latitude in an almost depth-independent manner, and interfaces with a nearly uniformly rotating radiative interior via the thin tachocline. How is that achieved? The matter was first seriously considered by Spiegel and Zahn (1992, see also \citet{eastachycline1972NASSP}).  
They, like I here, did not discuss the cause of differential rotation of the convection zone: it is evidently the result of a balance between the angular-momentum-transporting Reynolds stress, Maxwell stress and advection by large-scale meridional flow, a matter which is reviewed briefly by \citet{hanasogeetalISSI2013SSRv..tmp..100B} in this volume. Instead, recognizing that the global equilibration timescale of convection is no doubt much shorter than the dynamical timescales of processes operating beneath (even if they are related to the solar cycle), Spiegel and Zahn considered the convection-zone shear to be given, and ignored any back-reaction of the tachocline dynamics on the convection zone.  They then asked why the bulk of the fluid below rotates uniformly.

Spiegel and Zahn first established that although the radiative interior is very highly stratified, thermal diffusion is sufficiently efficacious to permit baroclinically driven flow (induced in a manner similar to that in my spin-down discussion of the previous section) to advect the required angular momentum essentially throughout the Sun in its lifetime. Therefore, the (imposed) differential rotation of the convection zone has to be insulated from the radiative interior to allow the latter to rotate uniformly. To achieve that, Spiegel and Zahn invoked a tachocline pervaded by turbulence generated by the tachocline shear itself, suppressed vertically by the stable stratification and therefore being layerwise two-dimensional. Moreover, they implicitly assumed the turbulence to be horizontally isotropic, notwithstanding the fact that the rms vorticity  of the turbulence is likely to be comparable with, or perhaps even less than, the angular velocity of the Sun, a situation which one would naturally expect to lead to substantial anisotropy. Thus they introduced a (turbulent) viscous stress tending to force the rotational flow towards being horizontally shear-free, and were thus able to achieve a steady state with a thin tachocline abutting a uniformly rotating interior, having angular velocity $\Omega_{\rm c}$. In the tachocline itself there was a flow similar to that in the Ekman layer in Figure 5: towards the axis of rotation -- i.e. poleward, on essentially horizontal surfaces -- in an equatorial region where the tachocline rotation exceeds that of the interior, and equatorward in polar regions. Spiegel and Zahn estimated that the two regions meet at latitude $42^{\rm o}$. That implied that $\Omega_{\rm c}$ is essentially the value of the photospheric angular velocity at that latitude, namely $0.90 \Omega_{\rm e}$, where $ \Omega_{\rm e}$ is the photospheric angular velocity at the equator. 

The study stimulated a potentially interesting numerical simulation by \citet{Miesch_mockrotatingturbulence2003ApJ...586..663M}, who set out to determine, amongst other matters, whether the shear turbulence would actually be likely to induce the rigidity produced by a scalar viscosity. He generated turbulence by imposing a grid of point sources of (gravity waves) in a stably stratified fluid under gravity. The grid was forced to rotate rigidly, and therefore the resulting mean flow inevitably tended towards rigid rotation too, in just the same way that gravity waves generated by wind over mountains exerts a drag on the wind (in the frame of reference in which the mountains are stationary). The conclusion that the turbulence leads to rigid rotation was therefore unjustified.  What would be much more revealing is to simulate a situation  having the sources move with the flow, so that no external torque is applied to the fluid.  

Spiegel and Zahn's study has no doubt introduced much of the pertinent dynamics of the tachocline. However, the details -- indeed even the basic principle -- have come under criticism. For example, it is unheard of that continuously maintained shear-generated turbulence can completely annul the shear that drives it. Indeed, an investigation by \citet{elliott1997A&A} demonstrated that even the magnitude of the turbulent stress generated would be insufficient for requirements. Moreover, in the observed natural environment, predominantly the Earth's (rotating) atmosphere, layerwise two-dimensional turbulence leads to augmentation, rather than 
suppression, of larger-scale shear, partly through angular-momentum transport by waves \citep{haynesetal1991JAtS...48..651H}. Such considerations led \citet{dogmem1998Nature} to conclude that the angular velocity of the radiative interior could never by kept uniform by fluid-dynamical processes alone, and that the interior must necessarily be held rigid by a large-scale magnetic field, presumably primordial. They outlined a nonlinear dynamical balance in what they regarded as the simplest model of the tachocline; it again led to a meridional flow, both geometrically and dynamically similar to that inferred by Spiegel and Zahn, associated with which is downflow from the convection zone in both polar and equatorial regions, and upflow between, near the latitude of zero shear, observed 
seismologically to be about $30^{\rm o}$ (see Figure 4), and implying that $\Omega_{\rm c}\simeq 0.93 \Omega_{\rm e}$ (cf. Figure 2). Advection of magnetic flux by the downflow counters the tendency for the field to expand by diffusion, yielding a steady-state balance which determines the thickness of the tachocline. Near the shear-free latitude the outflow might lift the primordial field into the convection zone, anchoring the angular velocity in the convection zone to that in the radiative interior, and possibly fuelling the 
magnetohydrodynamical processes responsible for the solar cycle \citep{stokednondynamos2014NJPh...16h3002B}. An obvious consequence of this picture is that in this region the shear might actually be quenched completely by the penetrating magnetic field, a feature that in principle is testable seismologically. It is interesting to note that the latitude of this region is the same as that at which sunspots first appear at the start of a solar cycle. It would be an unlikely coincidence if that had no dynamical significance. 

Of course, the potential role of a large-scale magnetic field holding the radiative interior rigid had been discussed before. I have already mentioned the work of  \citet{mestelweiss1987MNRAS.226..123M} and \citet{CharbonneauMacGregor1993ApJ...417..762C,CharbonneauMacGregor1993ApJ...403L..87C,CharbonneauMacGregor1993ApJ...417..762C} in connection with spin-down.  Additionally, \citet{RuedigerKitchatiniv1997AN....318..273R}  imagined the presence of such a field, without careful regard to the direct action on it of the differentially rotation convection zone, and proposed the tachocline to be just a simple Hartmann layer \citep[e.g.][]{Hartmann1937, shercliff1965tema.book.....S, paulrobertsmhdbook1967imhd.book.....R}  
as an explanation of why it is so thin. What Gough and McIntyre argued is that the magnetic field with a poloidal component must necessarily be present in the radiative zone, and they provided an albeit approximate analysis of how it is prevented from crossing (most of) the tachocline by gyroscopic pumping from the convection zone, thereby disenabling it from imparting on the radiative interior the latitudinal rotational shear in the convection zone. Some aspects of the analysis were subsequently supported by numerical simulations 
carried out by \citet{kitchatinovrudigertachocline2006A&A...453..329K} and \citet{rudigerkitchatinovtachocline2007NJPh....9..302R},  in particular the advection domination by the meridional flow, although the models differ in their details.  R\"udiger and Kitchatinov assumed a meridional flow much more rapid than the value derived by Gough and McIntyre, and produced a tachocline above an essentially 
uniformly rotating interior with a less intense field. An essential feature of both descriptions is the absence of Maxwell stress at the base of the convection zone. There are more complicated pictures in which the tachocline is  magnetized and turbulent \citep[e.g.][]{MEMtachocline2007sota.conf..183M,diamondetal2007sota.conf..213D}, a result of the nonlinear breakup of the flow resulting from the magnetorotational instability \citep{velikhov_MRI,chandrasekhar_MRI1960PNAS...46..253C,balbus-hawley_MRI1991ApJ...376..214B} or the Tayler instability \citep{tayler_instability1957PPSB...70...31T,tayler_instability_toroidal1973MNRAS.161..365T,markey-tayler_instability_poloidal1973MNRAS.163...77M} -- 
see also \citet{tayler-spruit_dynamo2002A&A...381..923S,arltsulerudiger2007A&A...461..295A,kitchatinov-rudiger_stability2007AN....328.1150K,rudigerkitchatinovtachocline2007NJPh....9..302R} --  and these are perhaps more representative of reality. 

Baroclinic meridional flow in the tachocline is an inevitable consequence of the basic steady-state dynamics of any model in which Maxwell stresses play only a minor role within the body of the tachocline itself. Therefore a prediction of the latitude of the confluence of the poleward and equatorward flows provides an easily testable consequence of the theory, because it determines $\Omega_{\rm c}$, which has been measured seismologically. That latitude in the Spiegel-Zahn theory is obtained from a linear equation, and is well determined once the latitudinal dependence of the turbulent stress-strain relation is specified (Spiegel and Zahn assume  it to be constant, without comment). By contrast, assuming a local linear stress-strain relation across the tachocline, with a coefficient of proportionality that is independent of latitude, yields $\Omega_{\rm c} \simeq 0.96 \Omega_{\rm e}$  \citep{Gough1985}.    The Gough-McIntyre theory is fundamentally nonlinear, and has not yet been worked out in sufficient detail for a prediction to be made. Some simpler toy models have been looked at too \citep[e.g.][]{garaud-guervilly_toy_tachocline_model2009ApJ...695..799G}, which at least shed some light on the potentially relevant processes: vertical shear, horizontal advection, geometry of Maxwell and Reynolds stresses.

One of the issues that needs to be addressed in the Gough-McIntyre picture is how the configuration could have been established. Numerical simulations by \citet{brunzahn2006A&A} and \citet{strugarekbrunzahn2011A&A...532A..34S,strugareketaltachocline2011AN....332..891S},  for example, have 
failed to reproduce an appropriately confined interior magnetic field, leading the authors to conclude that the 
theory must be wrong.    But the establishment depends critically on the dominant role of field advection, which could not be reproduced by the simulations because, as in all such computations, realistically small diffusion coefficients could not be achieved. Greater success has been achieved in a programme led by  Garaud, who 
was able to reduce diffusion by restricting attention to axisymmetric configurations (a restriction also adopted originally in the discussion by Gough and McIntyre). One of the consequent difficulties that she found initially   
is that any field penetrating the tachocline at the poles could not readily be advected equatorwards \citep[e.g.][]{garaudtachoclineI2002MNRAS}, leaving a large stress on the axis which locked the rotation of the convection zone and the radiative interior, even though an unlocked steady-state solution was subsequently shown to 
exist \citep{tswmemtach2011JFM}.  However, the most recent work \citep{Acavedoarraguin-garaud-wood2013MNRAS.434..720A} has succeeded for the first time to reproduce a steady two-dimensional 
numerical configuration consistent with the Gough-McIntyre theory, and provide a simple explanation of why previous numerical investigations had failed.

The assumption of axisymmetry, adopted originally for superficial simplicity, is perhaps also an extremely inhibiting impediment to the success of many of the the simulations. Provided other completing influences cannot dominate, an axisymmetric magnetic field, for example, whose axis of symmetry is initially nearly but 
not exactly aligned with the rotation axis, could perhaps be advected away from the rotation axis by the 
equatorward meridional flow in the tachocline, because it is on the magnetic axis that the field strength is greatest. The field would be left, maybe, with its axis of symmetry intersecting the tachocline at the latitude of the poleward-equatorward confluence: the latitude of zero shear. There it penetrates into the convection zone, where 
perhaps it suppresses the tachocline shear in a non-zero range of latitude.   A seismological test for such an outcome is currently being developed.   Indeed, it is not wholly out of the question that a similar process  is at least partially responsible for the obliquity of the orientation of large-scale magnetic fields in 
earlier-type stars \citep{DOG2012GApFD}.   More complicated asymmetric configurations can be envisaged. The outcome would be essentially a steady magnetic field configuration rotating with angular velocity $\Omega_{\rm c}$, which overall leads to a nonaxisymmetric Sun, even if the field itself, in its appropriate frame of reference, were axisymmetric.  Could this be the root of an explanation for the existence of active longitudes?   The asymmetry would rotate with the Sun, maintaining its phase.  However, long-term phase stability of the active longitudes is not obviously evident in the observations  \citep[e.g.][]{gyenge_etal_active_longitudes_2014SoPh..289..579G}.    Nevertheless, other indicators could be more telling:   it is interesting that the sector structure of the solar wind, which might mirror the magnetic asymmetry within the Sun,  has maintained its phase for at least four sunspot cycles prior to the mid seventies  \citep{svalgaard_wilcox_longtermsectorstructure1975SoPh...41..461S}.  Further, similar, analysis of solar-wind data up to the present day would therefore be very welcome, and could add (or subtract) credence to the hypothesis.

\begin{figure}
\centering
\includegraphics[width=0.85\linewidth]{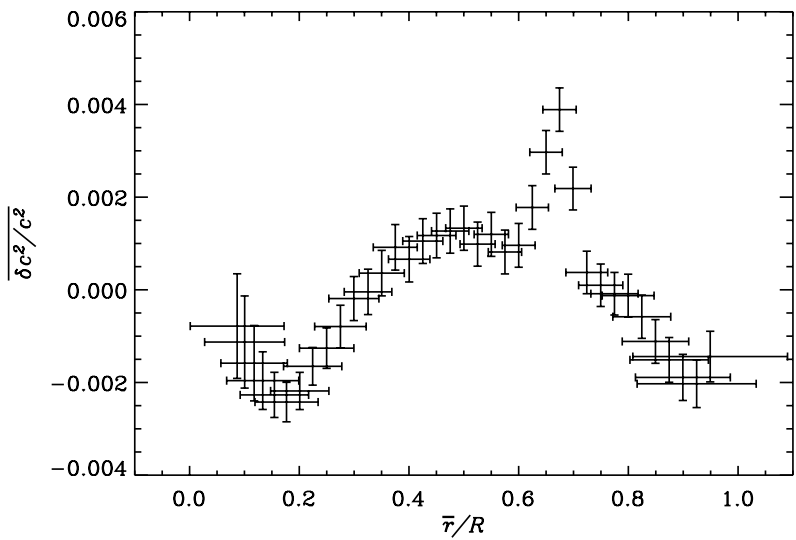}
\caption{Optimally localized  averages of relative differences between the squared sound speed  in the Sun and in Model S of \citet{jcdetal1996Sci}, computed by M. Takata from MDI 360-day data and plotted against the centres $\bar x = \bar r /R$ of the averaging kernels  $A(x;\bar x)$ --  which here resemble Gaussian functions --  and are defined by $\bar x = \int x A^2 {\rm d}x\,/ \int  A^2 {\rm d}x$.   The length of each horizontal bar is {\it twice} the spread $s$  of the corresponding averaging kernel, defined as $s=12\int(x-\bar x)^2A^2{\rm d}x$;  an averaging kernel $A$ that is well represented by a Gaussian function of variance $\Delta^2$ has spread approximately $1.7\Delta\simeq0.72FWHM$; were it to be a top-hat function, its spread would be the full width, which is why $s$ is so defined.  The vertical bars extend to $\pm$ one standard deviation of the errors, computed from the frequency errors quoted by the observers assuming them to be statistically independent with zero mean.  The sharp positive anomaly centred at ${\bar x} \simeq 0.67$  immediately beneath the convection zone is no doubt the consequence of chemical homogenization of the tachocline with the convection zone. The outward decline in ${\overline{ \delta c^2 / c^2}}$ in the convection zone is a result of having underestimated the seismic radius of the Sun.  The convex variation about ${\bar x} \simeq 0.15$ provides a hint of there having been some large-scale meridional flow in the core \citep[cf.][]{DOGAGK1990ASSL..159..327G}, which may also be responsible for the low equatorially biassed angular velocity evident in Figure 2.  The broad positive discrepancy in the radiative envelope is not understood.
} 
\label{fig6}
\end{figure}

Descriptions of the genre that I have discussed have all led to a ventilation timescale by the baroclinic meridional flow of order $10^6$y. That is long enough for thermal diffusion to establish a radiatively balanced thermal stratification, yet too short to be countered by microscopic diffusion, or gravitational settling of chemical species. Therefore the tachocline is chemically homogeneous with the convection zone, and thus has lower mean molecular mass than it would have had had gravitational settling been unopposed. Therefore the sound 
speed is higher than it would otherwise have been, creating the positive anomaly prominent in Figure 6. \citet{elliottdogtachthickness1999ApJ...516..475E} used that property to calibrate the thickness $\Delta$ of the tachocline\footnote{Here I adopt the original \citet{easjpztach1992A&A} definition of the tachocline: the gyroscopically pumped shear layer confined to only the stable region beneath the convection zone, despite the etymology of the appellation.}, yielding  $\Delta\simeq 2.0\times 10^{-2}R$, which is more precise (and smaller) than seismological measurements of shear, because hydrostatic stratification can be measured more precisely than rotation; however, the outcome depends on the reliability of the value of the diffusion coefficients required for evaluating the extent of gravitational settling, so the estimate may not be as accurate. 
Indeed, a yet unpublished investigation by Christensen-Dalsgaard and myself has established that the precise form of the tachocline anomaly cannot easily be reproduced by standard spherically symmetrical solar-structure theory, suggesting, perhaps, the presence of a degree of asphericity of the tachocline, in contradiction to an earlier finding of \citet{basuantia_tachoclineshape2001MNRAS.324..498B}. From an assessment of the horizontal balance of forces, it is inconceivable that the base of the tachocline is both steady and aspherical, although the transition between the stably stratified tachocline and the unstably stratified convection zone, aside from the inevitable convective buffeting, could be.  

Finally, a word about the overall shear near the base of the convection zone. Most seismological investigations of the tachocline have -- quite naturally, given its appellation -- considered only the shear itself, and have tried to characterize its thickness by fitting to the ill resolved inversions for $\Omega$  a chosen functional form with a quantifiable width \citep[e.g.][]{AGKtachocline1996ApJ}. Not surprisingly, the values found tended to exceed that determined from the sound-speed anomaly \citep{AGKtachocline1996ApJ,basu1997MNRAS.288..572B,antiaetal_tachocline1998MNRAS.298..543A,corbardetal_tachocline_thickness1998A&A...330.1149C,charbonneauetaltachocline1999ApJ}, because, in accord with the original definition, the true tachocline shear exists only in the stably stratified interior, although  \citet{corbardetal_tachocline_thickness1999A&A...344..696C} suggested subsequently that $\Delta$ might be 
as low as $0.01R$.   Almost all theoretical studies to date have ignored the reaction of the convection zone to 
the tachocline shear, which must penetrate to some degree into the convection zone, especially at high latitudes where vortex stretching is the greatest.  Of course, some upward penetration of the vertical shear must occur, and is indeed clearly visible in the seismological inferences (Figure 2), especially in the polar 
regions where vertical shear is resisted the most strongly by vortex stretching.  It is that penetration that has led to the conclusion by \citet{AGKtachocline1996ApJ,antiaetal_tachocline1998MNRAS.298..543A,charbonneauetaltachocline1999ApJ} and \citet{basuantiarotationvariation2003ApJ...585..553B} that the tachocline is thicker at the poles. A recent detailed helioseismic study of the shear, including its temporal variations, has been presented by \citet{antiabasu2011ApJ...735L..45A}.

\section{The temporally varying tachocline}
\label{sec:7}
The description of the tachocline outlined in the previous section is dominated by essentially laminar dynamics, superposed upon which there might be some weak small-scale turbulence providing additional, diffusive, transport. Moreover, the bulk of the tachocline is usually considered to be free of magnetic field, except in the upwelling region of near-zero shear. But there are more complicated situations that have been considered, and which suffer temporal variation on a timescale much shorter than the $10^6$-year ventilation time mentioned above. 

The most obvious time variation to consider is the buffeting by large-scale plumes from the convection zone. This is likely to cause the stability boundary -- i.e. the surface across which the (local) convective stability changes -- to undulate, in a manner similar to the tops (and bottoms) of terrestrial clouds (which, it should be pointed out, are not demarcated precisely by the visible vapour interface). Beneath that interface are tight 
ripples -- gravity waves predominantly with timescales and horizontal lengthscales comparable with those of the buffeting -- because the stable stratification is very much more intense than the unstable stratification above, with vertical lengthscales very much less than the horizontal scales. The group velocity is very nearly horizontal, so the ripples penetrate only a very short vertical distance before they dissipate (e.g. Gough, 1977). The undulations are usually called overshooting, but they do not necessarily induce as much mixing as is often presumed for causing the region of nearly adiabatic stratification to penetrate downwards and terminate in a 
yet sharper interface   \citep[e.g.][]{JCDmariorempelMJTovershoot2011MNRAS}). Instead they induce a smoothing of the horizontally averaged stratification (to which global seismology is sensitive). There is, 
however,  a small degree of material mixing resulting from minor disruptions to the interface. The gravity waves themselves also contribute to material transport by (nonlinear) \citet{taylor1953RSPSA.219..186T,taylor1954RSPSA.223..446T} dispersion, though on a much smaller vertical scale
 \citep[e.g.][]{press-rybicki_Taylordiffusion1981ApJ...248..751P,taylor-like_dispersion1992ApJ...401..196K}. 
It is likely that the pressure perturbations associated with the convective plumes are transmitted to the tachopause -- the base of the tachocline -- causing undulation in the boundary beneath.   A 
consequence of the smoothing of the horizontally averaged stratification near the base of the convection zone is a reduction in the magnitude of the (negative) leading constant in the asymptotic formula for the periods of high-order g modes of low degree \citep{ellis_g-mode_phase_constant1984LIACo..25..290E}, which could be detectable should the modes be observed.  At present there is no hope of a direct observation.  However, indirect methods, such as seeking g-mode-induced undulations in p modes, the dynamics of which I refrain from discussing here, may one day be successful.

The simple Gough-McIntyre description of the tachocline is magnetic-field free, except where it interfaces with the primordial field in the interior, and also near the almost shear-free region where the tachocline flow is upwelling. There is undoubtedly also a  magnetic field in the convection zone above, vacillating with the solar cycle with a characteristic 22-year period. If the temporally averaged field were strictly zero, the field would not penetrate far into the tachocline \citep[e.g.][]{pascalebpropagationt1999MNRAS}, which is why in their most elementary description the possible existence of such a field was ignored by Gough and McIntyre. But isn't it more likely that there is a randomness in the vacillation that leaves a nonzero residue? The residual field crossing the tachocline would be wound up by the tachocline shear. The situation would be ripe for Tayler-type  instability, as \citet{spruit1999A&A...349..189S,tayler-spruit_dynamo2002A&A...381..923S} has advocated \citep[see also][]{diamondetal2007sota.conf..213D}. It would lead to additional transport by the ensuing turbulence, and it would modify the relation between the tachocline thickness and the large-scale interior magnetic field \citep{MEMtachocline2007sota.conf..183M}. 
I think it is fair to say that this very complex subject is still ill understood. It remains an area of active research, attracting not only those who regard themselves as solar physicists. For more information I refer the reader especially to the book on the tachocline edited by \citet{hughesrosnerweisstach2007}, and to subsequent publications by the contributing authors.

It has even been proposed that the residual vacillating field leaking from the convection zone is alone sufficient to render the rotation of the radiative interior uniform (Forg\'acs-Dajke and Petrovay, 2000, 2002), without recourse to requiring the rigidity of a large-scale field. The idea seems to demand that the leaking field that is generated by dynamic action in the differentially rotating convection zone somehow takes on a rigid configuration. What must surely be the case is that the field takes on at least part of the convection-zone shear, which is then transmitted through the tachocline. So far as I am aware, not even a highly idealized model of the 
pertinent  dynamics has been fully investigated. However, if one imagines a toy model in which a periodically oscillating source of field with zero mean generated in a convection zone with high, turbulent, scalar magnetic diffusivity rotating at the observed differential rate, and diffusing into the radiative interior below, where the diffusivity is very much lower, then the temporally averaged Maxwell stress on the interior does not vanish, and is such as to generate differential rotation in the radiative envelope of the same form, although not necessarily with the same amplitude, as that in the convection zone. 
Solar-cycle-related time dependence is also not out of the question. 
 \citet{antiachitregoughmagneicrotation2013mnras}  have embarked upon a programme to study potential dynamical consequences of the seismologically determined variations of the angular velocity in the convection zone.  It is the only truly dynamical study that stems directly from helioseismological inference. 
The magnetic field required to produce the seismically observed angular-velocity variations was obtained under the assumption that the only azimuthal force is solely a Maxwell stress,  and in this initial study  only the azimuthal balance of forces was considered.  The analysis is therefore far from complete. However, a magnetic field was obtained which oscillates not only in the body of the convection zone but also in the tachocline. That result therefore reinforces the opinion  that the early, steady, tachocline studies were grossly oversimplified.

I conclude this section by mentioning the so-called tachocline oscillation, an apparently wave-like oscillation in the angular velocity near the equatorial plane immediately above and below the base of the convection zone, with a period of about 1.3 years. It was discovered seismologically by \citet{howeetaltachoclineoscillations2000Sci}, with opposite phase in the convection zone from that at the top of the radiative interior, thus exhibiting a maximum of the shear amplitude in the tachocline. Soon afterwards the oscillation almost disappeared \citep{howeetal_notachoclineoscillation2011JPhCS.271a2075H}; the eye of 
faith of an imaginative observer might perceive hints of its return \citep[see][]{Howe2009}, although the evidence is not statistically significant in any plausible sense.  What is the restoring force? Howe et al. suggested a magnetic field, which is quite plausible. Indeed, there is (unpublished) evidence that the oscillation extends more deeply than has been reported in the literature, with a further maximum in the shear at about $r=0.55R$, but with rather lower amplitude, as one would expect if the field does not increase with depth faster than $\sqrt \rho$.  It is interesting, although perhaps merely coincidental,  that the intensity of the vertical component of the field required to produce such an oscillation would need to be about 1T, which is similar to, 
yet rather less than, the rough {\it ab initio} estimate of the global field in the radiative zone.\footnote{and similar also to field intensities commonly discussed elsewhere in connection with the tachocline \citep[e.g.][]{tayler-spruit_dynamo2002A&A...381..923S,gilman-cally_tachocline2007sota.conf..243G}, although others \citep[e.g.][]{dogmem1998Nature,kitchatinov-rudiger_stability2007AN....328.1150K}  have entertained weaker fields.}  Howe and her colleagues (personal communication) consider the evidence for the deep penetration of the oscillation to be too insecure for them to report, but in trying to understand the workings of the Sun it is useful for theorists at least to bear in mind the possibility. What might drive the oscillation? One possibility is its influence on the anisotropic dissipation of gravity waves, a process which drives the terrestrial QBO, to which I  turn my attention later.

\section{Excitation of seismic modes}
\label{sec:8}
It is generally agreed that in the Sun acoustic oscillation modes are intrinsically stable, their energy being absorbed into the background configuration of the star by the combined action of appropriately phased heat and momentum transport by convection and, to a lesser degree, radiative heat transfer in the convectively stably stratified interior. The modes are generated stochastically by the turbulence in the upper layers of the convection zone, by both random impulses from the turbulent motion and by the associated fluctuations in 
buoyancy. There has been a sequence of more-and-more sophisticated analyses of the processes involved \citep[e.g.][]{stein1967SoPh....2..385S,goldreichkeeley1977ApJ...212..243G,balmforthgough1990excitationSoPh..128..161B,balmforth_excitation1992MNRAS.255..639B,belkacemetal_excitation2009A&A...494..191B,chaplietal_modeamplitudes2005MNRAS.360..859C}.    They derive initially from the work of \citet{batchelorhomogeneousturbulence1953} and \citet{lighthill_aerodynamic_sound_generation_I_1952RSPSA.211..564L,lighthill_aerodynamic_sound_generation_II_1954RSPSA.222....1L}, who considered respectively 
the stochastic excitation of a (general) simple  harmonic oscillator and the mechanism of turbulent acoustic wave generation by momentum transfer --  also of \citet{DOG1965WHOI, DOG1977Nice,unno_conv_puls1967PASJ...19..140U,xiong1989A&A...209..126X,gabriel_t-d_mlt1996BASI...24..233G}, who addressed the role of convection in determining the linear growth and damping  rates of stellar oscillations \citep[see also][]{dupretetal_excitation2005A&A...435..927D,dupretetal_excitation2006ESASP.624E..97D,dupretetal_amplitudes_etc2009A&A...506...57D}. The absolute intensity of the nonlinear excitation is extremely difficult to quantify, because it is extremely sensitive to details of the turbulence, which are not well  
defined by the mixing-length formulations that are adopted to represent the solar convection zone \citep{DOG1977Nice, DOGDavos2002ESASP.508..577G}. Therefore the absolute amplitudes of the seismic waves are very uncertain. Nevertheless, an appropriate combination of the uncertain factors in the theory can be adjusted at least to harmonize with seismic observations; that leads to a functional form of amplitude against 
frequency -- a true prediction of the theory -- which is in reasonable accord with observation \citep{houdeknjbjcddog1999A&A...351..582H,chaplietal_modeamplitudes2005MNRAS.360..859C}. It is interesting, and perhaps somewhat surprising, that, once the adjustment has been made, the remaining weakest part of the theory, at least when applied to the Sun, then appears to be the prediction of the linear damping rates of 
the modes; that inference was obtained by replacing the theoretical damping rates by observed acoustic power-spectrum line widths, and finding  substantial amelioration  \citep{chaplietal_modeamplitudes2005MNRAS.360..859C}.

Further work on the convection-pulsation interaction is therefore called for, not merely to make minor improvements to derivations of mode amplitudes in the Sun, but, more importantly, to improve our understanding of the underlying dynamics, both for its own sake and for application to other stars for which the gross solar adjustment of the theory may no longer be appropriate. 

Observations of solar-like observations in other stars add new data with which to compare the theory. Particularly useful are stars rather different from the Sun. An observation of the mode amplitudes in the (coolish) more evolved star $\xi$ Hydrae, for example, has been well predicted  \citep{HG_xsi_Hydrae2002MNRAS.336L..65H}, although subsequent direct estimates of the damping rates \citep{stelloetal_xsiHydrae_linewidths2006A&A...448..709S} were not in accord with the theory, suggesting, perhaps, the presence of cancelling errors in the theoretical growth and excitation rates. Before jumping to that conclusion, however, one must recognize that the damping rates were obtained from the observations by equating them with the line widths in the power spectrum, which is a correct procedure only if it is known that the background state of the star is truly invariant over the duration of the observations \citep{batchelorhomogeneousturbulence1953}; otherwise non-dissipative phase (frequency) wandering can broaden the spectral lines and thereby falsify the conclusion. Indeed, subsequent analysis of high-quality data from the space missions CoRoT and Kepler \citep{huberetal_redgiantlinewidths_etc2010ApJ...723.1607H,baudinetal_red-giant_amplitudes_etc2011A&A...535C...1B}  yielded values in accord with the theoretical predictions.

The situation is much worse with the hot star Procyon, whose oscillation amplitudes are grossly overestimated   \citep{houdeknjbjcddog1999A&A...351..582H,matthewsetal_np_procyon_osc2004Natur.430...51M,matthewsetal_erratum2004Natur.430..921M}. In this case there must be something fundamentally wrong with how one describes either how the turbulence properties scale with variations of stellar structure, or, worse still, the manner in which they combine to quantify the excitation; alternatively, there might be some missing ingredient in the calculation of the damping rates, such as wave scattering or some process unaccounted for in the time-dependent convection theory.

\section{On the dynamics of gravity modes}
\label{sec:9}
In an attempt to resolve the solar neutrino problem before helioseismology, \citet{dilkegough1972Natur.240..262D} suggested that  intermittent nonlinear breakdown of a grave unstable gravity mode would trigger finite-amplitude convection that almost completely mixes the core. That suggestion is now known not to be correct, because the form of the depression of the sound-speed in the central regions of the Sun, evident in Figure 1, incontrovertibly demonstrates the existence of a substantial gradient of helium abundance \citep{DOGAGK1990ASSL..159..327G}. Moreover, an important analysis by \citet{wojtek1982AcA....32..147D} of triad coupling to a resonating pair of otherwise damped daughter modes suggested that it is unlikely that the principal unstable mode would ever achieve an amplitude high enough to overcome the barrier to the nonlinear onset of convection. Nevertheless, the dynamical processes that were thought to operate are not necessarily irrelevant today, and can perhaps be profitably rehearsed. 
The driving of grave solar g modes is principally by the modulation of the nuclear reactions, a mechanism understood first by \citet{eddingtoninternalconstitution1926ics}, and now called the $\epsilon$ mechanism.  Extreme sensitivity of the nuclear reaction rates to temperature causes (local) temperature excesses to be augmented in the face of (global) losses by radiative diffusion. Key to the process in cool stars like the Sun,  which are powered by the relatively insensitive p-p reaction chain, is the destruction of nuclear balance by the oscillations, which vary on the timescale of an hour, much shorter than the $10^5$-year nuclear equilibration time, exposing the variation of the energy generated by the temperature-sensitive $^3$He--$^3$He, and the $^3$He--$^4$He, nuclear reactions \citep[cf.][]{LedouxSauvenier-Goffin1950ApJ...111..611L}.
Also important is that the g-mode-depressing convection zone is deep compared with those in hotter stars, preventing the mode from achieving so great an amplitude in the outer stellar envelope as to suffer sufficient  damping to overcome the nuclear driving. Consistent calculations by \citet{jcdfwwddog1974MNRAS.169..429C}, \citet{shibahashiosakiunno1975PASJ...27..401S}, \citet{bouryetal_gmode_stability1975A&A....41..279B} and \citet{saio1980ApJ...240..685S} confirmed the original estimate that some g modes were indeed unstable. 

The analysis by  \citet{wojtek1982AcA....32..147D} is particularly interesting dynamically, because the outcome is superficially counterintuitive. It is well known that a pair of daughter gravity modes can sap energy from their parent, provided that they resonate precisely enough. What is less well appreciated are the conditions required for that process to be severely debilitating. 
When I was a student at the Program of Geophysical Fluid Dynamics (GFD) at the Woods Hole Oceanographic Institution in 1965 my attention was caught by the daily water being heated for coffee in an aluminium pan on a flat electric hob. The bottom of the pan was rounded with age, and the pan rocked on the heated plate. If the quantity of water was right, there was a resonance between the rocking frequency and the frequency of surface gravity modes in the water, and the differently located bursts of heat through the bottom of the pan as different parts of it touched the hot plate caused the oscillation to be sustained. Eddington's $\epsilon$ mechanism was operating at home. The oscillation amplitude was not tiny, yet surely the intrinsic dissipation -- measured by the inverse Reynolds number, for example -- was much greater than it is in the Sun. So shouldn't gravity modes in the Sun attain even greater amplitudes? The answer is negative: oddly enough, it is actually because the propensity for dissipation in the GFD pot was so high that the oscillation could be sustained. To understand that one must first realize that to be an effective energy drain the difference between the frequencies of the two daughters must match the frequency of the parent within an inverse intrinsic growth time, a condition which 
in general becomes harder and harder as the thermal diffusion coefficient, and hence the nonadiabaticity, diminishes. This requires finding daughter pairs amongst modes with higher and higher characteristic wave numbers, which actually leads to higher and higher damping rates, notwithstanding the smaller diffusion coefficient, therefore lower and lower filial amplitudes and consequently an even lesser and lesser ability of the daughters to extract energy from their parent. Only if the dissipation rate (diffusion coefficient) is low enough can the daughters attain high-enough amplitudes for them to be able to limit the growth of the parent effectively. 

Dziembowski assumed the background stratification of the Sun would not be changing with time. He found that it was not possible always to find daughter pairs sufficiently close to resonance, yet with adequately low damping rates to permit them to grow to amplitudes high enough to severely inhibit the parent. So he carried out a statistical calculation to estimate the expected amplitude of the parent, and found the amplitude to be a mere 10 cm s$^{-1} $ in the early stages of the Sun's evolution on the main sequence \citep{dziembowski1983SoPh...82..259D}. That is surely insufficient to cause serious disruption to the structure of 
the core. Nevertheless, it appeared to permit accidentally much higher amplitudes. What is the possible outcome? That question is answerable once it is realized that the Sun is evolving, and that resonance with particular daughters cannot be maintained.  In fact the passage from one resonating pair to another as the Sun evolves typically occurs on a timescale shorter than the growth time of the daughters. Moreover, it turns out that strict formal 
phase coherence over the time it takes for the parent to reach its limiting amplitude is not required: the parent's phase drifts as the Sun evolves, and as new energy-sapping daughters are encountered their phases are established by the triad interaction itself. Therefore, as reported briefly by \citet{Jordinson-DOD_g-mode_coupling2000ASPC..203..390J}, the limiting amplitude of the parent, as its frequency drifts amongst the family of resonances, is simply the expectation value calculated by Dziembowski.    A simplifying assumption of the original calculation was that interaction with only a single daughter pair was considered; in practice, 
however, there are interactions at any time with an entire hierarchy, limiting the amplitude of the parent yet more severely. However,  in general it is the gravest pair who sap by far the most energy, and it is likely that the extra loss to all the other daughter pairs is actually quite small, leaving Dziembowski's estimate more-or-less intact. 

It appears from this discussion that the structure of the Sun, even the temperature-sensitive $^8{\rm B}$ neutrino production, is unlikely to have been modified seriously by gravity modes.  Indeed, an end to this matter has been called by many astrophysicists who have grappled with neutrinos.  However, anyone interested in dynamics might still ask whether there is any small remnant consequence. There has for a long time \citep[cf.][]{DOGAGK1990ASSL..159..327G,dogagktt1995SoPh..157....1G} been a hint, albeit rather weak, that some degree of material redistribution, possibly laminar, has taken place in the inhomogeneous energy-generating core:  this is evinced by the generally negative slope of $\overline{\delta c^2/c^2}$ inwards of $r\simeq 0.2 R$, visible in Figure 6,  which appears to imply a radial gradient in the spherically averaged $Y$ 
that is somewhat lower than theoretical expectation, and which,  both in the very early days of helioseismology  and also more recently \citep{basuetal_core2009ApJ...699.1403B}, 
has been recognized as being suggestive of mild mixing.   Moreover, a general redistribution of angular momentum within the core from an initially uniformly rotating state could make the core appear to be rotating more slowly, as is suggested by Figure 2, because on the whole the angular velocity near the edge of the core, especially near the equatorial plane, where  rotational splitting is most prone to be affected, is likely to be diminished. Could these hints be manifestations of a soliton in the core, controlled either by gravity-wave-like dynamics, or by some more slowly operating dynamics that might be influencing reported variations in the neutrino luminosity \citep[e.g.][]{sturrock_rmodes2008ApJ...688L..53S}?  A possibly simpler, but less interesting, solution might be that mixing  in the early stages of the main-sequence evolution was much more extensive than it is now,  because the angular velocity (and especially its deceleration)  was much greater then than it is today, possibly causing adequate baroclinicity to drive a laminar circulation through the core when the impedance by chemical inhomogeneity was quite small.

Could there have been significant Taylor dispersion by daughter g modes? These are matters that appear not to have been addressed. One might note that the grave g modes have similar amplitudes both in the core and near the surface, and that Dziembowski's 10 cm s$^{-1}$ estimate is high enough to be observed at the surface today. However, the structure of the Sun today differs substantially from its early main-sequence state, and some calculations suggest that no g mode is now unstable \citep{bouryetal_gmode_stability1975A&A....41..279B,shibahashiosakiunno1975PASJ...27..401S}, although the matter is a delicate balance between driving and damping processes, leaving open the possibility that some g modes are today intrinsically unstable \citep{noelsetal1975MSRSL...8..317N,saio1980ApJ...240..685S}.  There have been various claims of direct detection, but these are not universally accepted.  I think it is fair to say that there is yet no 
convincing evidence for their existence at observable amplitudes \citep{Phoebus_quest_2010A&ARv..18..197A}.  Finding gravity modes would be extremely important for seismology; as has long been appreciated  \citep[e.g.][]{dog_pleinsfeaux1978pfsl.conf...81G,GOLFmission1995SoPh..162...61G}, the additional seismological resolving power that the g modes  would provide would be quite substantial. Are there modes that are unstable, or are they all intrinsically stable and excited stochastically as pressure fluctuations associated with the undulating tachocline or by the same pressure fluctuations that excite acoustic modes in the outer layers of the convection zone? Estimates of the latter are summarized by \citet{Phoebus_quest_2010A&ARv..18..197A}.  They vary quite widely, although amplitudes are typically lower than the observational claims.  A great deal of uncertainty remains.

\section{Gravity-wave transport}
\label{sec:10}
In an early discussion of gravity-wave transport between core and convection zone  \citep{DOG1977Nice} it was estimated that grave modes of the kind that might be self-excited in the core to amplitudes comparable with those discussed in the previous section are too weak to carry a substantial amount of heat, and too weak also to transfer a substantial proportion of the Sun's angular momentum in its lifetime. Moreover, stable gravity waves that resonate in space and time with the undulating lower boundary of the convection zone penetrate only about 1 km into the radiative interior. A more recent estimate of the latter \citep{DOGDavos2002ESASP.508..577G}  
suggests that penetration is about 100 times deeper, but that is still too shallow to have a material impact on the overall structure. It was estimated that modes generated off-resonance would have too low an amplitude to transport a substantial amount of angular momentum. It was also pointed out that any group of gravity waves that do penetrate must dissipate asymmetrically, and are expected to enhance, rather than suppress, differential rotation. 

Subsequently \citet{press1981ApJ...245..286P} and \citet{garcialopezspruit_gwaves_1991ApJ...377..268G} estimated the wave amplitudes by balancing inertially generated pressure fluctuations near the base of the convection zone against corresponding resonant gravity-wave fluctuations in the radiative zone beneath. As with the estimates of acoustic wave generation \citep[e.g.][]{DOG1977Nice}, there is substantial uncertainty arising from uncertainties in the properties of the convective turbulence. 

An important consequence of the enhancement of rotational shear by gravity-wave transport was illustrated with a pioneering experiment carried out by \citet{plumbmcewan1978JAtS...35.1827P}, who demonstrated that the necessarily anisotropic wave dissipation can induce global oscillations of the shear, with characteristic period inversely proportional to the fourth power of the amplitude of the gravity waves (and proportional to the sixth power of the horizontal phase speed). 
The experiment was carried out to confirm a theory of \citet{lindzenholton1968JAtS...25.1095L} and \citet{holtonlindzen1972JAtS...29.1076H} explaining the mechanism of the terrestrial QBO. It is of some interest  to summarize the basic process here, because there is misinformation in the literature. 

Low-amplitude gravity waves generated with a broad spectrum of frequencies at an essentially horizontal surface propagate upwards (in the case of the Earth's atmosphere) into the sheared stratosphere. 
The large-scale flow in the stratosphere is essentially horizontal, its variation with time and horizontal distance being much weaker than the frequency and horizontal wavenumber of the waves (and was therefore ignored in discussing the linearized local wave dynamics). Frequency and horizontal wavenumber in an inertial frame of reference are therefore conserved. Given that the gravity-wave dispersion relation is approximately $\omega^2 = N^2k_{\rm h}^2/(k_{\rm v}^2+k_{\rm h}^2)$, where $\omega$ is the wave frequency in the local frame of the fluid, $N$  is the buoyancy (Brunt-V\"a\"is\"ala) frequency of the stratification, and $k_{\rm h}$ and  $k_{\rm v}$ are respectively the horizontal and vertical components of the wave number, it follows that the horizontal component of the group velocity, and the (angular) momentum flux, are in the same direction as the horizontal phase velocity\footnote{The vertical component of the group velocity is directed oppositely to the phase  velocity.}. 
Also prograde waves, having negatively Doppler-shifted frequencies $\omega$, have higher $k_{\rm v}$ than do the corresponding retrograde waves, and therefore dissipate faster,  causing the  momentum that is imparted on the mean flow to be preferentially in the prograde direction. If prograde and retrograde waves are excited to the same amplitude, the shear is therefore at first enhanced, but further from the excitation layer,  where the remnant prograde waves have lower amplitude  as a result of earlier higher dissipation, the direction of the wave stress acting on the mean flow is reversed. 

An initially static atmosphere into which gravity waves are propagating is (then) susceptible to shear enhancement.  If initially there is no mean shear, a random fluctuation in the symmetry of the wave excitation will first produce an initial weak shear which is then amplified by the process described above. The increased shear near the excitation layer eventually loses momentum to the excitation layer, and the whole pattern descends, producing a temporal oscillation in the mean horizontal flow at any given altitude. Evidently the amplitude and period of that oscillation must depend on the excitation rate and the dissipation rate, the latter, in view of the dispersion relation, depending on the horizontal phase velocity. The existence of the oscillation requires the fluid not to be too diffusive, for otherwise the oscillation is damped. 

Is there an analogous phenomenon in the Sun?  Likely amplitude estimates  lead to a characteristic period of about a month \citep{DOGDavos2002ESASP.508..577G}, in a very thin layer within the tachocline into which the waves penetrate before being dissipated. It is most unlikely that any significance can be attributed to the coincidence of that period with the characteristic large-eddy turnover time at the base of the convection zone or with the rotation period of the Sun. In fact there may be no such coincidence, because the amplitudes of the 
waves generated, and hence the period of the oscillation in the shear generated, are so uncertain. A characteristic period even a hundred times longer is not out of the question. Indeed,  \citet{dograpporteur_QBO-solarcycle1998SSRv...85..141G} wondered whether Plumb and McEwan's (non-magnetic) mechanism could be responsible for the 22-year solar cycle, leaving the magnetic field to play only a secondary role in the dynamics.  The process could also induce meridional flow \citep{frittsvadasandreassen_gwaves1998A&A...333..343F}, which might conceivably provide a significant transport mechanism for chemical species, and possibly explain the Li deficiency in the Sun's photosphere.    An example of a rather different claim comes from an investigation by \citet{kumartalonzahn1999ApJ...520..859K}, who estimated an enormously greater gravity-wave amplitude for waves that penetrate more deeply\footnote{as had both \citet{kumarquataert1997ApJ...475L.143K} and \citet{zahntalonmatias_gwavetransport1997A&A...322..320Z} earlier, in the mistaken belief that angular momentum transport by the waves would be in the reverse sense and would thus supply the reason for the observed uniform rotation of the radiative envelope.}. 
They too commented  that Plumb and McEwan's (non-magnetic) mechanism might be responsible for the 22-year solar cycle. In a subsequent investigation, Talon, Kumar and Zahn (2002) argued that shear instability would soften the oscillations. The effect of the instability was modelled with a scalar viscosity which increased the QBO-like oscillation period to about 300y, and the spectrum of the waves had enough power at low order to permit angular momentum to be redistributed throughout the radiative interior, again supposed to establish an almost uniform rotation profile. How the waves would suppress the imprint of the differential rotation in the convection zone discussed by \citet{easjpztach1992A&A} remained unexplained. 

The simplest discussions of this process are based on the action of an almost monochromatic group of waves \citep[e.g.][]{kimmacgregor2001ApJ...556L.117K,kimmacgregor2003ApJ...588..645K}.  A more realistic broad spectrum of waves such as would be expected in the Earth's atmosphere \citep{baldwinetal2001RvGeo..39..179B}  and the Sun can lead to much more complicated behaviour, as \citet{mathisetal_2008SoPh..251..101M} portend. From a series of numerical simulations, \citet{rogersmacgregorglatzmaier2008MNRAS.387..616R}, \citet{rogersmacgregor2010MNRAS.401..191R,rogersmacgregor2011MNRAS.410..946R} and \citet{macgregorrogers2011SoPh..270..417M}
have concluded that simple linearized wave dynamics is inadequate for computing the mean Reynolds stress because (nonlinear) wave-wave interactions are important at the amplitudes 
they expected to be generated near the lower boundary of the solar convection zone.  
They argued that
oscillatory shear flow of the QBO type is much less likely to be produced --  although rigid rotation is not to be expected either, as \citet[e.g.][]{MEM1994seit.conf..293M,MEMtachocline2003safd.book..111M} has often 
stressed.  One might wonder whether in the numerical simulations excessive diffusion requires unrealistically large wave amplitudes, as is possibly the case also of convective motion \citep{ToomreThompsonISSI2015}, and that the 
quasi-linear dynamics of the kind described by \citet{plumbmcewan1978JAtS...35.1827P} really is a good guide.  We await with interest the further, more realistic, simulations that are promised.


\bibliographystyle{aps-nameyear}      
\bibliography{ISSIreferences}   

\end{document}